\documentclass[twocolumn]{aastex631}

\usepackage{bm}
\usepackage{xcolor}
\usepackage[normalem]{ulem}

\defcitealias{Mandel22}{M20}
\defcitealias{Thorp21}{T21}
\defcitealias{Avelino19}{Avelino}

\begin{document}

\title{
Relative intrinsic scatter in hierarchical Type Ia supernova siblings analyses: Application to SNe~2021hpr, 1997bq $\&$ 2008fv in NGC~3147}

\author{Sam~M.~Ward}
\affiliation{Institute of Astronomy and Kavli Insititute for Cosmology, Madingley Road, Cambridge, CB3 0HA, UK}

\author{Stephen~Thorp}
\affiliation{Institute of Astronomy and Kavli Insititute for Cosmology, Madingley Road, Cambridge, CB3 0HA, UK}
\affiliation{The Oskar Klein Centre, Department of Physics, Stockholm University, AlbaNova University Centre, SE 106 91 Stockholm, Sweden}

\author{Kaisey~S.~Mandel}
\affiliation{Institute of Astronomy and Kavli Insititute for Cosmology, Madingley Road, Cambridge, CB3 0HA, UK}
\affiliation{Statistical Laboratory, DPMMS, University of Cambridge, Wilberforce Road, Cambridge, CB3 0WB, UK}
\affiliation{The Alan Turing Institute, Euston Road, London, NW1 2DB, UK}

\author{Suhail~Dhawan}
\affiliation{Institute of Astronomy and Kavli Insititute for Cosmology, Madingley Road, Cambridge, CB3 0HA, UK}

\author{David~O.~Jones}
\affiliation{Department of Astronomy and Astrophysics, University of California, Santa Cruz, CA 95064, USA}

\author{Kirsty~Taggart}
\affiliation{Department of Astronomy and Astrophysics, University of California, Santa Cruz, CA 95064, USA}

\author{Ryan~J.~Foley}
\affiliation{Department of Astronomy and Astrophysics, University of California, Santa Cruz, CA 95064, USA}

\author{Gautham~Narayan}
\affiliation{University of Illinois at Urbana-Champaign, 1003 W. Green St., IL 61801, USA}
\affiliation{Centre for Astrophysical Surveys, National Centre for Supercomputing Applications, Urbana, IL 61801, USA}

\author{Kenneth~C.~Chambers}
\affiliation{Institute for Astronomy, University of Hawaii, 2680 Woodlawn Drive, Honolulu, HI 96822, USA}

\author{David~A.~Coulter}
\affiliation{Department of Astronomy and Astrophysics, University of California, Santa Cruz, CA 95064, USA}

\author{Kyle~W.~Davis}
\affiliation{Department of Astronomy and Astrophysics, University of California, Santa Cruz, CA 95064, USA}

\author{Thomas~de~Boer}
\affiliation{Institute for Astronomy, University of Hawaii, 2680 Woodlawn Drive, Honolulu, HI 96822, USA}

\author{Kaylee~de~Soto}
\affiliation{Department of Astronomy \& Astrophysics, The Pennsylvania State University, University Park, PA 16802, USA}
\affiliation{Institute for Computational \& Data Sciences, The Pennsylvania State University, University Park, PA, USA}

\author{Nicholas~Earl}
\affiliation{University of Illinois at Urbana-Champaign, 1003 W. Green St., IL 61801, USA}

\author{Alex~Gagliano}
\affiliation{University of Illinois at Urbana-Champaign, 1003 W. Green St., IL 61801, USA}
\affiliation{Centre for Astrophysical Surveys, National Centre for Supercomputing Applications, Urbana, IL 61801, USA}

\author{Hua~Gao}
\affiliation{Institute for Astronomy, University of Hawaii, 2680 Woodlawn Drive, Honolulu, HI 96822, USA}

\author{Jens~Hjorth}
\affiliation{DARK, Niels Bohr Institute, University of Copenhagen, Jagtvej 128, 2200 Copenhagen, Denmark}

\author{Mark~E.~Huber}
\affiliation{Institute for Astronomy, University of Hawaii, 2680 Woodlawn Drive, Honolulu, HI 96822, USA}

\author{Luca~Izzo}
\affiliation{DARK, Niels Bohr Institute, University of Copenhagen, Jagtvej 128, 2200 Copenhagen, Denmark}

\author{Danial~Langeroodi}
\affiliation{DARK, Niels Bohr Institute, University of Copenhagen, Jagtvej 128, 2200 Copenhagen, Denmark}

\author{Eugene~A.~Magnier}
\affiliation{Institute for Astronomy, University of Hawaii, 2680 Woodlawn Drive, Honolulu, HI 96822, USA}

\author{Peter~McGill}
\affiliation{Department of Astronomy and Astrophysics, University of California, Santa Cruz, CA 95064, USA}

\author{Armin~Rest}
\affiliation{Department of Physics and Astronomy, The Johns Hopkins University, Baltimore, MD 21218, USA}
\affiliation{Space Telescope Science Institute, Baltimore, MD 21218, USA}

\author{C\'{e}sar~Rojas-Bravo}
\affiliation{Department of Astronomy and Astrophysics, University of California, Santa Cruz, CA 95064, USA}

\author{Rados{\l}aw~Wojtak}
\affiliation{DARK, Niels Bohr Institute, University of Copenhagen, Jagtvej 128, 2200 Copenhagen, Denmark}

\collaboration{25}{(Young Supernova Experiment)}



\begin{abstract}
We present Young Supernova Experiment $grizy$ photometry of SN~2021hpr, the third Type Ia supernova sibling to explode in the Cepheid calibrator galaxy, NGC~3147. 
Siblings are useful for improving SN-host distance estimates, and investigating the contributions towards the SN~Ia intrinsic scatter (post-standardisation residual scatter in distance estimates). 
We thus develop a principled 
Bayesian framework for analyzing SN~Ia siblings. 
At its core is the cosmology-independent relative intrinsic scatter parameter, $\sigma_{\rm{Rel}}$: 
the dispersion of siblings distance estimates relative to one another within a galaxy. 
It quantifies the contribution towards the total intrinsic scatter, $\sigma_0$, from within-galaxy variations about the siblings' common properties. 
It also affects the combined-distance uncertainty. 
We present analytic formulae for computing a $\sigma_{\rm{Rel}}$-posterior from individual siblings distances (estimated using any SN-model). 
Applying a newly trained \textsc{BayeSN} model, we fit the light curves of each sibling in NGC~3147 individually, to yield consistent distance estimates. 
However, the wide $\sigma_{\rm{Rel}}$-posterior 
means $\sigma_{\rm{Rel}}\approx \sigma_0$ is not ruled out. 
We thus combine the distances by marginalizing over $\sigma_{\rm{Rel}}$ 
with an informative prior: $\sigma_{\rm{Rel}}\sim U(0,\sigma_0)$. 
Simultaneously fitting the trio's light curves improves constraints on distance, \textit{and} each sibling's individual dust parameters, compared to individual fits. 
Higher correlation also tightens dust parameter constraints. 
Therefore, $\sigma_{\rm{Rel}}$-marginalization yields robust estimates of siblings distances for cosmology, and dust parameters for siblings-host correlation studies. 
Incorporating NGC~3147's Cepheid-distance yields
$H_0=78.4\pm 6.5\,\text{km\,s}^{-1}\text{\,Mpc}^{-1}$. 
Our work motivates analyses of homogeneous siblings samples, to constrain $\sigma_{\rm{Rel}}$, and its SN-model dependence.
\end{abstract}

\keywords{
Type Ia supernovae(1728) --- Distance indicators(394) --- Interstellar dust extinction(837) --- Astrostatistics(1882) --- Cosmology(343) --- Light curves(918)
}

\section{Introduction}
Type Ia supernovae (SNe~Ia) are standardisable candles, used to measure luminosity distances and constrain cosmological parameters~\citep{Riess98:lambda, Perlmutter99, Abbott19:dessn, Freedman21, Stahl21, Brout22, Jones22, Riess22}. The increasing number of observed SNe~Ia~\citep[e.g.][]{Riess99:lc, Jha06:lc, Hicken09:lc, Hicken12, Krisciunas17, Foley18:found, Hounsell18,Abbott19:dessn, Ivezic19,Jones19, Jones21, Rose21, Scolnic21, Brout22, Jones22}, means cosmological analyses will be dominated by systematic uncertainties, rather than statistical uncertainties. Key to reducing these systematics is to understand the origin of the $\approx 0.1$~mag total intrinsic scatter, $\sigma_0$, i.e. the post-standardisation residual scatter in SN distance estimates compared to a best-fit cosmology~\citep{Scolnic19, Brout21}.  

SN~Ia siblings -- SNe Ia that occur in the same host galaxy -- are valuable for investigating standardisation systematics~\citep{Elias81,Hamuy91,Stritzinger11, Brown15, Gall18, Burns20, Scolnic20,Scolnic21,Graham22, Kelsey23}. Siblings share common properties, like distance, redshift, and host galaxy stellar mass. Consequently, the relative dispersion of siblings distance estimates has no contribution from a variation of these common properties. This mitigation of systematic uncertainties can yield insights into the origin of the total intrinsic scatter. Siblings are also useful for improving distance estimates to the host galaxy, by combining individual siblings distances.

In this work, we develop a principled Bayesian framework for robustly analyzing SN Ia siblings. We do this by introducing the `relative intrinsic scatter' parameter, $\sigma_{\rm{Rel}}$. This is the intrinsic scatter of individual siblings distance estimates relative to one another within a galaxy. It is constrained without assuming any cosmology, and quantifies the contribution towards the total intrinsic scatter, $\sigma_0$, from within-galaxy variations about the siblings' common properties.

We model $\sigma_{\rm{Rel}}$ to analyze a unique trio of SN~Ia siblings in NGC~3147 --- the only Cepheid calibrator galaxy to host three SNe~Ia. We advance on previous state-of-the-art single-galaxy siblings studies. Like \cite{Hoogendam22,Barna23}, we find strong consistency between individual siblings distance estimates, but we compute a $\sigma_{\rm{Rel}}$-posterior to assess the probability that $\sigma_{\rm{Rel}}>0$. Further, we marginalize over $\sigma_{\rm{Rel}}$ with an informative prior, $\sigma_{\rm{Rel}}\sim U(0,\sigma_0)$, to robustly quantify the uncertainty on a combined distance estimate. For the first time, we simultaneously fit siblings light curves to yield improved constraints on distance and host galaxy dust parameters, compared to individual fits. Similar to \cite{Biswas22}, 
we study the SN Ia color-luminosity relation in a cosmology-independent fashion, but we constrain a physically-motivated common host galaxy dust law parameter $R_V$. Like \cite{Gallego-Cano22}, we use a single siblings calibrator calibrator galaxy to estimate the Hubble constant. Our inference is dominated by the Cepheid-distance measurement error; nonetheless, for the first time, we robustly propagate the siblings' combined uncertainty into a cosmological inference, by marginalizing over $\sigma_{\rm{Rel}}$. 

We present our correlated intrinsic scatter model in \S\ref{S:RelativeScatter}. We then apply these concepts to fit NGC~3147's siblings. The data and model are described in \S\ref{S:ObsandData} and \S\ref{S:BayeSN}, respectively; this includes new Pan-STARRS-1 $grizy$ photometry of SN~2021hpr from the Young Supernova Experiment~\citep{Jones21}, and a new `W22' version of \textsc{BayeSN}~(\citealt{Mandel22}; Appendix~\ref{AppendixNewModel}). We perform our analysis in \S\ref{S:Analysis}, and conclude in \S\ref{S:Conclusions}.

\section{Relative Intrinsic Scatter}
\label{S:RelativeScatter}
\subsection{Correlated Intrinsic Scatter Model}
\label{S:IntrinsicScatterModel}
We decompose the distance estimate error terms originating from the total intrinsic scatter\footnote{The total $\delta M_s$ parameter is the achromatic offset, with variance $\sigma_0^2$, which contributes towards the total variance in the Hubble residuals, in addition to the redshift-based distance errors and the photometric distance measurement errors.}, $\delta M_s$, into common and relative components. The achromatic magnitude offset common to all siblings in a galaxy is $\delta M_{\rm{Common}}$, and the relative achromatic offsets -- specific to each SN sibling -- are $\delta M_{\rm{Rel},s}$.
\begin{equation}
\label{eq:dMtotal}
    \delta M_{s} = \delta M_{\rm{Common}} + \delta M_{\rm{Rel},s}
\end{equation}
The population distributions of each component are:
\begin{eqnarray}
\delta M_{\rm{Common}} &\sim& \mathcal{N}(0,\sigma^2_{\rm{Common}}), \\
    \delta M_{\rm{Rel},s} &\sim& \mathcal{N}(0,\sigma^2_{\rm{Rel}}).
\end{eqnarray}
Their sum must have a dispersion equal to the total intrinsic scatter, $\sigma_0$:
\begin{equation}
\label{eq:VardMs}
\text{Var}[\delta M_s]=\sigma_0^2 =\sigma^2_{\rm{Common}}+\sigma^2_{\rm{Rel}}.
\end{equation}
The common intrinsic scatter, $\sigma_{\rm{Common}}$, is thus the contribution towards $\sigma_0$ from population variations of the siblings' common properties. Meanwhile, the relative intrinsic scatter, $\sigma_{\rm{Rel}}$, is the contribution towards $\sigma_0$ from within-galaxy variations about the siblings' common properties. This simple model can be applied in hierarchical Bayesian settings to robustly analyze siblings, and combine individual siblings distance estimates.

The uncertainty in a joint siblings distance estimate
has a contribution, $\text{Var}[\overline{\delta M}]$, from the intrinsic scatter components:
\begin{equation}
\label{eq:VardMbar}
    \text{Var}[\overline{\delta M}] = \sigma_0^2-\sigma_{\rm{Rel}}^2\Big(1-\frac{1}{N_{\rm{Siblings}}}\Big),
\end{equation}
where $\overline{\delta M}$ is the siblings' sample mean $\delta M_s$ . Therefore, if the siblings are assumed to be perfectly correlated, $\sigma_{\rm{Rel}}=0$, the distance uncertainty contribution is $\sigma_0$ (it is maximised). But when the siblings are assumed to be perfectly uncorrelated, $\sigma_{\rm{Rel}}=\sigma_0$, the distance uncertainty contribution is minimised: $\sigma_0/\sqrt{N_{\rm{Siblings}}}$. Computing the precision-weighted average of individual siblings distance estimates is equivalent to adopting this uncorrelated assumption, $\sigma_{\rm{Rel}}=\sigma_0$. Consequently, the joint distance uncertainty may be underestimated if in fact there is some correlation, i.e. $\sigma_{\rm{Rel}}<\sigma_0$.

\subsection{$\sigma_{\rm{Rel}}$ Prior Knowledge}
\label{S:SigmaRelPrior}

The size of the relative scatter is uncertain in the literature. On the one hand, \cite{Scolnic20} and \cite{Scolnic21} analyzed 8 and 12 siblings galaxies, respectively, using SALT2~\citep{Guy07, Guy10}, and results indicated $\sigma_{\rm{Rel}}$ is large and comparable to the total intrinsic scatter, $\sigma_0$. This implies the siblings distance estimates are highly uncorrelated. On the other hand, \cite{Burns20} constrain the dispersion of \textsc{SNooPy}~\citep{Burns11} distance differences between sibling pairs; after removing fast decliners, and SNe observed with the Neil Gehrels Swift Observatory, they use 11 siblings galaxies to estimate a 0.03~mag 95\% upper bound on their dispersion hyperparameter. This is sub-dominant compared to the total intrinsic scatter in the Hubble diagram (typically $\sim 0.1-0.15$~mag), which implies the siblings distance estimates are highly correlated. It is unclear then how $\sigma_{\rm{Rel}}$ compares to $\sigma_0$, and how this depends on the SN sample properties, and the SN model used to estimate distances. 

A priori, we expect $\sigma_{\rm{Rel}}\leq \sigma_0$. This is because siblings in each galaxy share some common properties, which may otherwise vary in the population and directly contribute towards the total intrinsic scatter. These common properties include: distance, redshift, peculiar velocity, and common host galaxy properties. The common host properties include both the global host galaxy properties, e.g. host galaxy stellar mass, and the siblings' mean local host galaxy properties, e.g. mean SFR, mean metallicity, mean stellar age etc. The reported correlations of SN~Ia Hubble residuals with global and/or local host galaxy properties~\citep{Sullivan03,Kelly10, Sullivan10,DAndrea11, Childress13,Rigault13, Pan14, Uddin17, Jones18:local,Roman18, Rigault20,Rose20,Smith20, Uddin20, Brout21, Johansson21, Kelsey21, Ponder21,Popovic21, Thorp21,Briday22, Meldorf22,Thorp22, Wiseman22}, indicates $\sigma_{\rm{Common}}>0$, or equivalently, $\sigma_{\rm{Rel}}<\sigma_0$. 

Moreover, $\sigma_{\rm{Rel}}$ is constrained without using `redshift-based distances': distances obtained by inputting estimates of cosmological redshift into a cosmological model. Therefore, $\sigma_{\rm{Rel}}$ has no error contribution from estimating cosmological redshift, and assuming a cosmology; however, these two sources may contribute towards $\sigma_0$ if these additional errors -- if any -- are not already encompassed by the redshift-based distance uncertainties. 

The contrast of $\sigma_{\rm{Rel}}$ versus $\sigma_0$ thus indicates whether it is within-galaxy variations, or the population variation of the siblings' common properties, that dominates $\sigma_0$. We conclude that $0\leq\sigma_{\rm{Rel}}\leq\sigma_0$ is expected, and $\sigma_{\rm{Rel}}$ can be marginalized over with an informative flat prior: 
\begin{equation}
\sigma_{\rm{Rel}} \sim U(0,\sigma_0).
\end{equation}
\textbf{This is justified because the contributions towards $\sigma_{\rm{Rel}}$ are a subset of those that contribute towards $\sigma_0$.}

\subsection{Constraining $\sigma_{\rm{Rel}}$}
\label{S:ConstrainingSigmaRel}
We present analytic formulae for computing a $\sigma_{\rm{Rel}}$-posterior using individual siblings distance estimates. In a single galaxy, we adopt a simple normal-normal hierarchical Bayesian model, as described in Chapter 5.4 of \citet{Gelman13}. Re-writing in our notation, we have $N_{\rm{Siblings}}$ individual distance estimates, $\hat{\mu}_s$, each with a measurement uncertainty, $\hat{\sigma}_{\rm{fit},\,s}$, from fitting each set of siblings light curves, $\mathcal{D}_s$. They are estimates of the distance plus the common and relative achromatic offsets, $\mu+\delta M_{\rm{Common}}+\delta M_{\rm{Rel},s}$; therefore, there is a dispersion of the latent variables, which comes from the relative scatter: $\sigma_{\rm{Rel}}$. This is written as:
\begin{equation}
    \hat{\mu}_s \sim \mathcal{N}(\mu+\delta M_{\rm{Common}},\sigma^2_{\rm{Rel}}+\hat{\sigma}^2_{\rm{fit},\,s}).
\end{equation}
Using eq. 5.21 from \cite{Gelman13} to marginalize over the distance and common offset, $\mu+\delta M_{\rm{Common}}$, and with a $\sigma_{\rm{Rel}}$-prior, $p(\sigma_{\rm{Rel}})$, we can compute an un-normalised relative scatter posterior:
\begin{equation}
\label{eq:indepsigma0posterior1}
    p(\sigma_{\rm{Rel}} | \bm{\mathcal{D}}) \propto \frac{p(\sigma_{\rm{Rel}})\prod_{s=1}^{N_{\rm{Siblings}}} \mathcal{N}(\hat{\mu}_s | \hat{\mu}_w, \hat{\sigma}^2_{\rm{fit},\,s}+\sigma_{\rm{Rel}}^2)  }{\sqrt{\sum_{s=1}^{N_{\rm{Siblings}}}w_s}  },
\end{equation}
\begin{equation}
\label{eq:indepsigma0posterior2}
    w_s = \frac{1}{\hat{\sigma}^2_{\rm{fit},\,s} + \sigma_{\rm{Rel}}^2}\,\,\,\,\,;\,\,\,\,\,
    \hat{\mu}_w = \frac{\sum_{s=1}^{N_{\rm{Siblings}}} w_s \hat{\mu}_s  }{\sum_{s=1}^{N_{\rm{Siblings}}} w_s},
\end{equation}
where $\bm{\mathcal{D}}=\{\mathcal{D}_s\}^{N_{\rm{Siblings}}}_{s=1}$. We note the single-galaxy $\sigma_{\rm{Rel}}$ likelihoods are conditionally independent~\citep{Bishop06}. Therefore, a multi-galaxy $\sigma_{\rm{Rel}}$-posterior is computed simply by multiplying the prior by the product of likelihoods in Eq.~\ref{eq:indepsigma0posterior1}. We use the siblings in NGC~3147 to compute a single-galaxy $\sigma_{\rm{Rel}}$ posterior in \S\ref{S:IndepFits}.

\begin{figure}
    \includegraphics[width=1\linewidth]{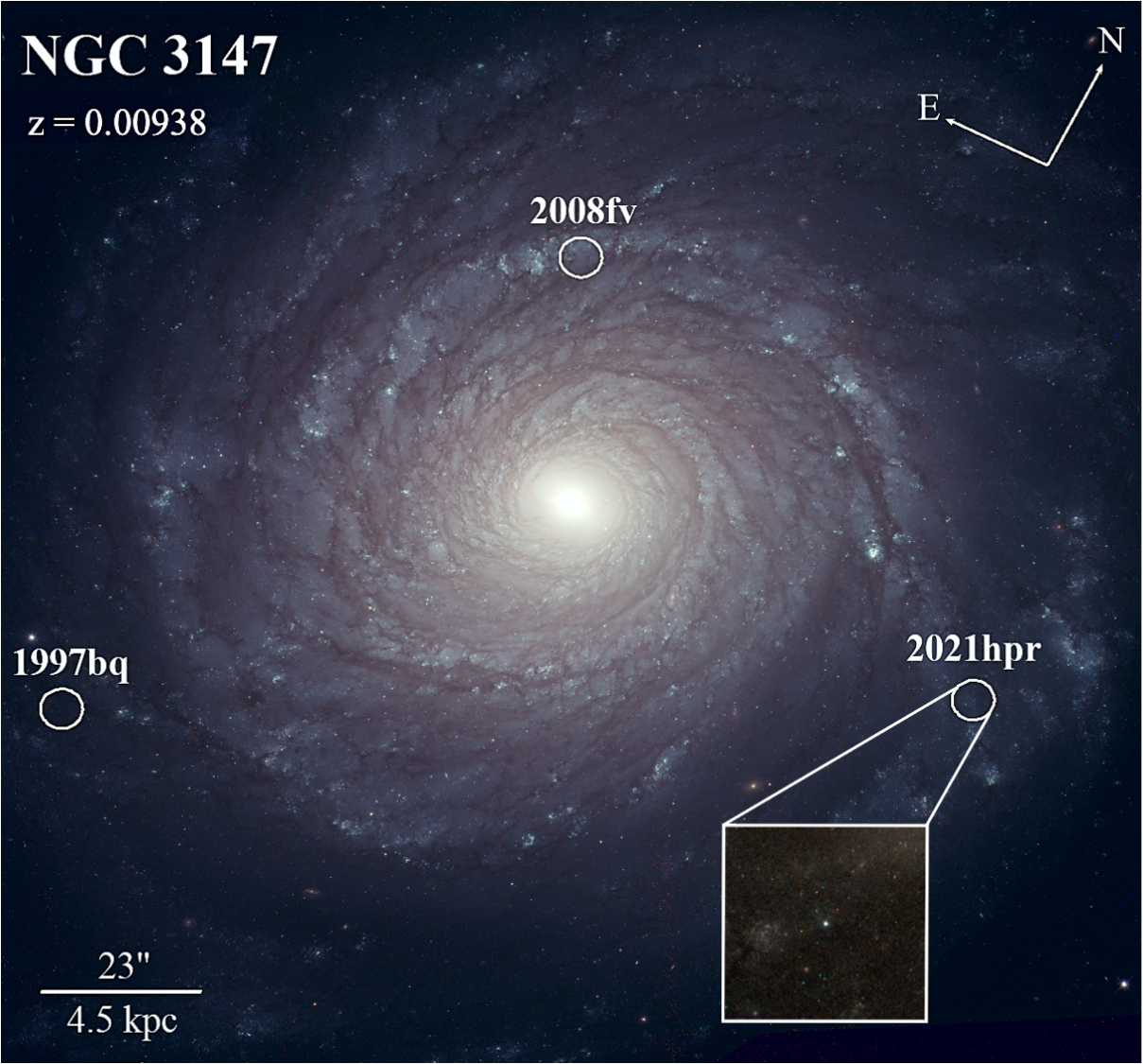}
    \caption{
    \textit{Hubble Space Telescope} image of NGC~3147 ($z_{\rm{Helio}}=0.00938$), with the SN~Ia trio's locations marked, and the length scale in the bottom left. Overlayed is a $15.8'' \times 15.8''$ (3~kpc on a side) inset region of SN~2021hpr from 30$^{\rm{th}}$ December 2021 ($\approx 9$~months after SN detection).
    }
    \label{fig:GalaxyPic}
\end{figure}

\begin{deluxetable*}{c c c c c c c c c c c c }
\label{tab:21hprphotometry}
\tablecaption{New SN~2021hpr $grizy$ photometry, observed with Pan-STARRS-1 as part of the Young Supernova Experiment~\citep{Jones21}\,\tablenotemark{a}.}
\tablehead{
MJD & Phase\tablenotemark{b} & $g$\tablenotemark{c} & $\sigma_g$ & $r$ & $\sigma_r$ & $i$ & $\sigma_i$ & $z$ & $\sigma_z$ & $y$ & $\sigma_y$
}
\startdata
59312.38 & -8.99 & 15.055 & 0.006 & 15.109 & 0.005 & 15.499 & 0.007 & 15.464 & 0.008 & 15.567 & 0.015 \\
59313.37 & -8.01 & 14.889 & 0.004 & 14.913 & 0.004 & 15.283 & 0.004 & 15.332 & 0.005 & 15.409 & 0.012 \\
59314.29 & -7.10 & 14.710 & 0.003 & 14.785 & 0.003 & 15.123 & 0.004 & 15.223 & 0.005 & 15.297 & 0.009 \\
59315.24 & -6.16 & - & - & - & - & 15.013 & 0.004 & 15.092 & 0.005 & 15.230 & 0.009 \\
59315.25 & -6.15 & 14.592 & 0.004 & 14.659 & 0.003 & - & - & - & - & - & - \\
 & & & & & \multicolumn{2}{c}{...} \\
\enddata
\vspace{-0.1cm}
\tablenotetext{a}{
The full data will be made available online upon publication.
}
\vspace{-0.2cm}
\tablenotetext{b}{Phase is computed with $(T_{B;\,\rm{max}}, z_{\rm{Helio}}) = (59321.46~\rm{d},0.00938)$ using Eq.~\ref{eq:phasecomp}.}
\vspace{-0.2cm}
\tablenotetext{c}{All photometry in magnitudes.}
\end{deluxetable*}

\begin{deluxetable}{r r r c}
\label{tab:data}
\tablecaption{Dataset summary of NGC 3147's SN~Ia trio.}
\tablehead{
SN
& Alias
& Reference 
& Passbands\tablenotemark{a}
}
\startdata
2021hpr& 21hpr&  Table~\ref{tab:21hprphotometry} & $grizy$ \\
2008fv& 08fv  & \cite{Scolnic21} & $BVR$ \\
1997bq& 97bq& \cite{Scolnic21} & $BVRI$ \\
\enddata
\vspace{-0.1cm}
\tablenotetext{a}{Only the passbands that can be fitted with \textsc{BayeSN} are shown, which excludes the SN~1997bq $U$-band data.}
\end{deluxetable}

\vspace{-1cm}
\section{Data}
\label{S:ObsandData}

We apply these concepts to analyze light curves of NGC~3147's trio of SN~Ia siblings. We present new Young Supernova Experiment Pan-STARRS-1 $grizy$ photometry of SN~2021hpr in Table~\ref{tab:21hprphotometry}~\citep{Chambers16}. First detected on 2$^{\rm{nd}}$ April 2021~\citep{Itagaki21}, SN~2021hpr is the third spectroscopically confirmed normal Type Ia supernova in NGC 3147 ($z=0.00938$), which is a high-stellar-mass Cepheid calibrator galaxy~\citep{Rahman12,Yim16,Sorai19}. The data were reduced with {\tt Photpipe} \citep{Rest05:photpipe}, which has been used for numerous photometric reductions of YSE data (e.g.~\citealp{Kilpatrick21,Tinyanont21,Dimitriadis22,Gagliano22,Jacobson-Galan22:21gno,Jacobson-Galan22:20tlf,Terreran22}). The {\tt YSE-PZ} software was used for a transient discovery alert of SN~2021hpr, and for data collation, management, and visualisation~\citep{Coulter22:ysepz, Coulter23}.
We present a spectrum of SN~2021hpr in Appendix~\ref{S:Appendix2021hprSpectrum}, which indicates it is a normal SN~Ia. 
Recently, SN~2021hpr was studied using optical photometry and spectral observations~\citep{Zhang22, Barna23, Lim23}.

For the other two siblings, SNe~1997bq and 2008fv, we use the most up-to-date and well-calibrated Pantheon+\footnote{\url{https://github.com/PantheonPlusSH0ES/DataRelease.git}} light curves~(\citealt{Jha06:lc,Tsvetkov10,Scolnic21}; summary in Table~\ref{tab:data}).

Fig.~\ref{fig:GalaxyPic} displays a \textit{Hubble Space Telescope} image of NGC~3147, with the trio's locations marked. The image\footnote{\url{https://dx.doi.org/10.17909/k2kh-6h52}} was constructed from stacked images produced by and retrieved from the Barbara A.\ Mikulski Archive for Space Telescopes, obtained from Programmes GO--15145 (PI Riess) and SNAP--16691 (PI Foley) between 29$^{\rm{th}}$ October 2017 and 30$^{\rm{th}}$ December 2021.  The SN~2021hpr inset was created using the Trilogy software~\citep{Coe12} using data from 30$^{\rm{th}}$ December 2021 obtained through Programme SNAP--16691.

\section{Modeling}
\label{S:BayeSN}
We use \textsc{BayeSN} to fit each sibling's light curves individually, or all the siblings' light curves simultaneously. \textsc{BayeSN} is an optical-to-near-infrared hierarchical Bayesian model of the SN~Ia SEDs~\citep{Thorp21, Mandel22}. It uniquely enables us to fit the trio's optical-to-near-infrared light curve data to coherently estimate distance and dust parameters. 

\subsection{New \textsc{BayeSN} Model: W22}
\label{S:NewModel}
The trio's data are in $BVRI$ and $grizy$ passbands, whereas previous iterations of \textsc{BayeSN} were trained either on $BVRIYJH$ or $griz$, but not both. Therefore, we retrain a new and more robust \textsc{BayeSN} model (hereafter `W22') simultaneously on optical-NIR $BgVrizYJH$ (0.35--1.8$\mu$m) data. The training sample combines the Foundation DR1~\citep{Foley18:found, Jones19} and \cite{Avelino19} samples, so comprises
236 SNe~Ia in host galaxies that are $\approx 70\%$ high-mass ($\log_{10}(\rm{M}_{*}/\rm{M}_{\odot}) > 10$). This sample does not include the siblings trio analyzed in this work. Population hyperparameters learned in W22 model training include the global dust law shape, $R_V=2.659$, and the total intrinsic scatter, $\sigma_0=0.094$~mag. Appendix~\ref{AppendixNewModel} provides further details on this new model. 

\subsection{Fitting Procedures}
\label{S:fittingprocedures}
To correct for Milky Way extinction, we use the \cite{Fitzpatrick99} law, adopting $R_{V;\,\rm{MW}}=3.1$, and a reddening estimate $E(B-V)_{\rm{MW}}=0.024$~mag from \cite{Scolnic21}. To model host galaxy dust, we fit for each sibling's dust law shape parameter, $R_V^s$, using a flat prior: $R_V^s\sim U(1,6)$. The lower bound is based on the Rayleigh scattering limit $R_V\approx 1.2$~\citep{Draine03}, and the upper bound is motivated by observational results of lines of sight in the Milky Way~\citep{Fitzpatrick99, Schlafly16}. 

The time of $B$-band maximum brightness, $T_{B;\,\rm{max}}$, which defines rest-frame phase via:
\begin{equation}
\label{eq:phasecomp}
    t = \frac{T_{\rm{MJD}}-T_{B;\,\rm{max}}}{1+z_{\rm{Helio}}},
\end{equation}
is fitted for in an individual fit, and thereafter frozen at the posterior mean time. Data outside the model phase range, $-10<t<40$~d, and data with $\rm{SNR}<3$, are removed from the fit.

\subsection{Joint Siblings Fit Modeling Assumptions}
\label{S:ModellingAssumptions}

In the new joint fits\footnote{\url{https://github.com/bayesn/bayesn-public/tree/siblings/BayeSNmodel/stan_files}}, we fit all light curves of siblings in a single galaxy simultaneously, while applying a common distance constraint; this yields a posterior on a single distance hyperparameter.

\textbf{To jointly fit the siblings, we must make an assumption about their correlation.} While the total intrinsic scatter is learned in \textsc{BayeSN} model training, $\sigma_0=0.094$~mag, the $\sigma_{\rm{Rel}}$ value is unknown. Therefore, we can fit under three $\delta M$ modeling assumptions shown in Table.~\ref{tab:IntrinsicScatterModellingAssumptions}. 
We can assume the siblings are perfectly uncorrelated, perfectly correlated, or fit for and marginalize over $\sigma_{\rm{Rel}}$ while imposing an informative prior: $\sigma_{\rm{Rel}}\sim U(0,\sigma_0)$. Respectively, these are the $\delta M$-Uncorrelated, $\delta M$-Common, or $\delta M$-Mixed assumptions (Table.~\ref{tab:IntrinsicScatterModellingAssumptions}). \textbf{Our default assumption is to marginalize over $\sigma_{\rm{Rel}}$.}

\begin{deluxetable}{l l l}
\label{tab:IntrinsicScatterModellingAssumptions}
\tablecaption{Intrinsic scatter modeling assumptions for hierarchically analyzing siblings.}
\tablehead{
Assumption & $\sigma_{\rm{Rel}}$-prior & $\mu$-uncertainty\,\tablenotemark{a}
}
\startdata
$\delta M$-Uncorrelated & $\sigma_{\rm{Rel}}=\sigma_0$ & $\sigma_0/\sqrt{N_{\rm{Siblings}}}$\\
     $\delta M$-Mixed & $\sigma_{\rm{Rel}}\sim U(0,\sigma_0)$ & $(\sigma_0/\sqrt{N_{\rm{Siblings}}},\sigma_0)$\tablenotemark{b} \\
      $\delta M$-Common & $\sigma_{\rm{Rel}}=0$ & $\sigma_0$\\
\enddata
\tablenotetext{a}{The contribution towards the joint siblings distance uncertainty originating from the total intrinsic scatter.}
\vspace{-0.2cm}
\tablenotetext{b}{Marginalizing over $\sigma_{\rm{Rel}}$ means a posterior on $\sigma_{\rm{Rel}}$ is computed, which weights the common distance uncertainty accordingly. The uncertainty contribution given some value of $\sigma_{\rm{Rel}}$ is shown in Eq.~\ref{eq:VardMbar}.}

\end{deluxetable}

\begin{deluxetable*}{l l c | c c c c | c}
\label{tab:intialindepfits}
\tablecaption{Posterior summaries of SN parameters from individual fits, and the joint common-$\mu$ fit, to the siblings' light curves.}
\tablehead{Fit Type / Summary & Dataset(s) & Bands & $\mu$ (mag)& $A^s_{V}$ (mag)&$\theta_s$& $R^s_{V}$\tablenotemark{a} & $\hat{\sigma}_{\rm{fit},s}$(mag)\,\tablenotemark{b}
}
\startdata
&
21hpr & $grizy$
&
$33.131 \pm 0.119$
 & 
$0.266 \pm 0.069$
 & 
$-0.566 \pm 0.097$
 & 
$2.801 \pm 1.249$ & $0.071$
\\
Individual
&
97bq & $BVRI$
&
$33.127 \pm 0.157$
 & 
$0.404 \pm 0.132$
 & 
$\,\,\,\,0.134 \pm 0.194$
 & 
$2.758 \pm 1.210$ & $0.128$
\\
&
08fv & $BVR$
&
$33.256 \pm 0.219$
 & 
$0.441 \pm 0.200$
 & 
$-0.607 \pm 0.176$
 & 
$<3.112(4.927)$ & $0.198$
\\
\tableline
&
21hpr& $grizy$
&
& 
$0.244 \pm 0.061$
 & 
$-0.562 \pm 0.094$
 & 
$2.572 \pm 1.174$
\\
Common-$\mu$
&
97bq& $BVRI$
&
$33.169 \pm 0.107$
 & 
$0.379 \pm 0.082$
 & 
$\,\,\,\,\,0.143 \pm 0.178$
 & 
$2.546 \pm 0.949$ & -
\\
&
08fv& $BVR$
&
 & 
$0.537 \pm 0.097$
 & 
$-0.580 \pm 0.158$
 & 
 $2.860 \pm 0.754$
\\
\enddata
\vspace{-0.1cm}
\tablenotetext{a}{The 68 (95)\% quantiles are quoted for posterior peaking at zero. The dust law shape priors are $R_V^s \sim U(1,6)$.}
\vspace{-0.2cm}
\tablenotemark{b}{The fitting uncertainties, or `measurement errors', on the individual siblings distance estimates, computed using Eq.~\ref{eq:estimatesigmatlam}.}
\end{deluxetable*}

\vspace{-1cm}
\section{Analysis}
\label{S:Analysis}

\subsection{Individual Fits}
\label{S:IndepFits}

\begin{figure}
    \centering
    \includegraphics[width=1\linewidth]{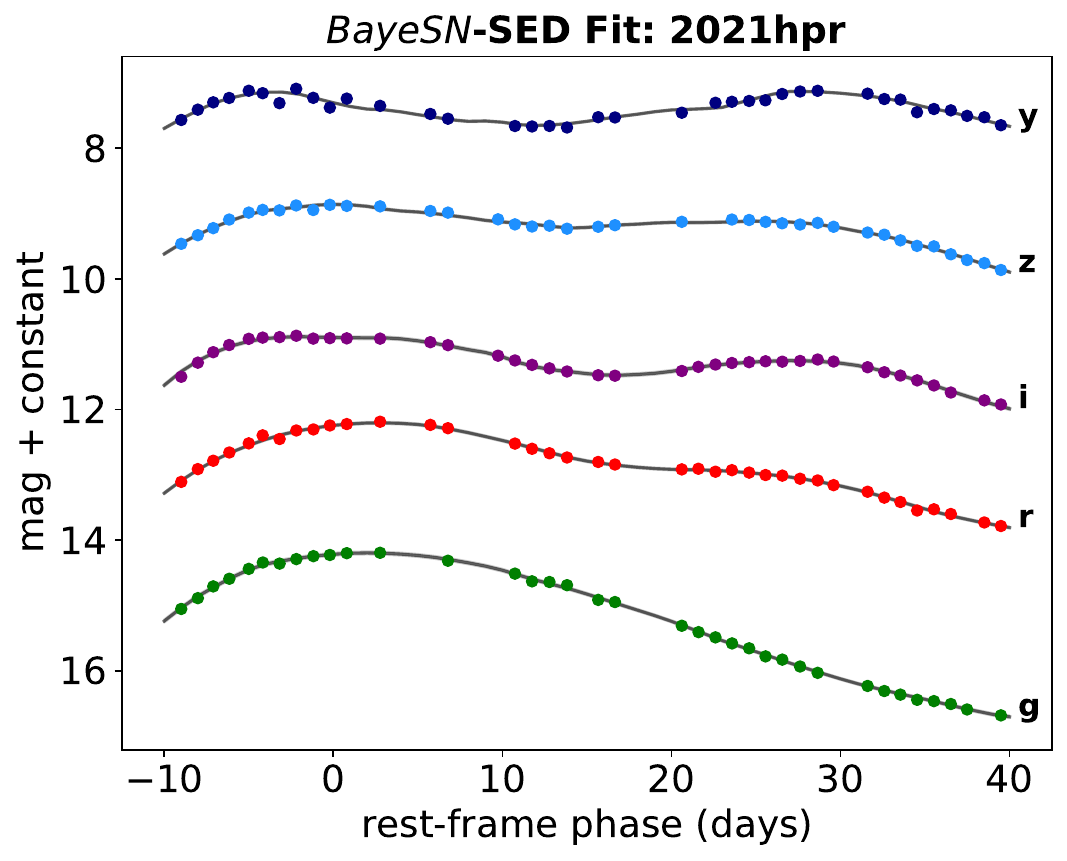}
    \includegraphics[width=1\linewidth]{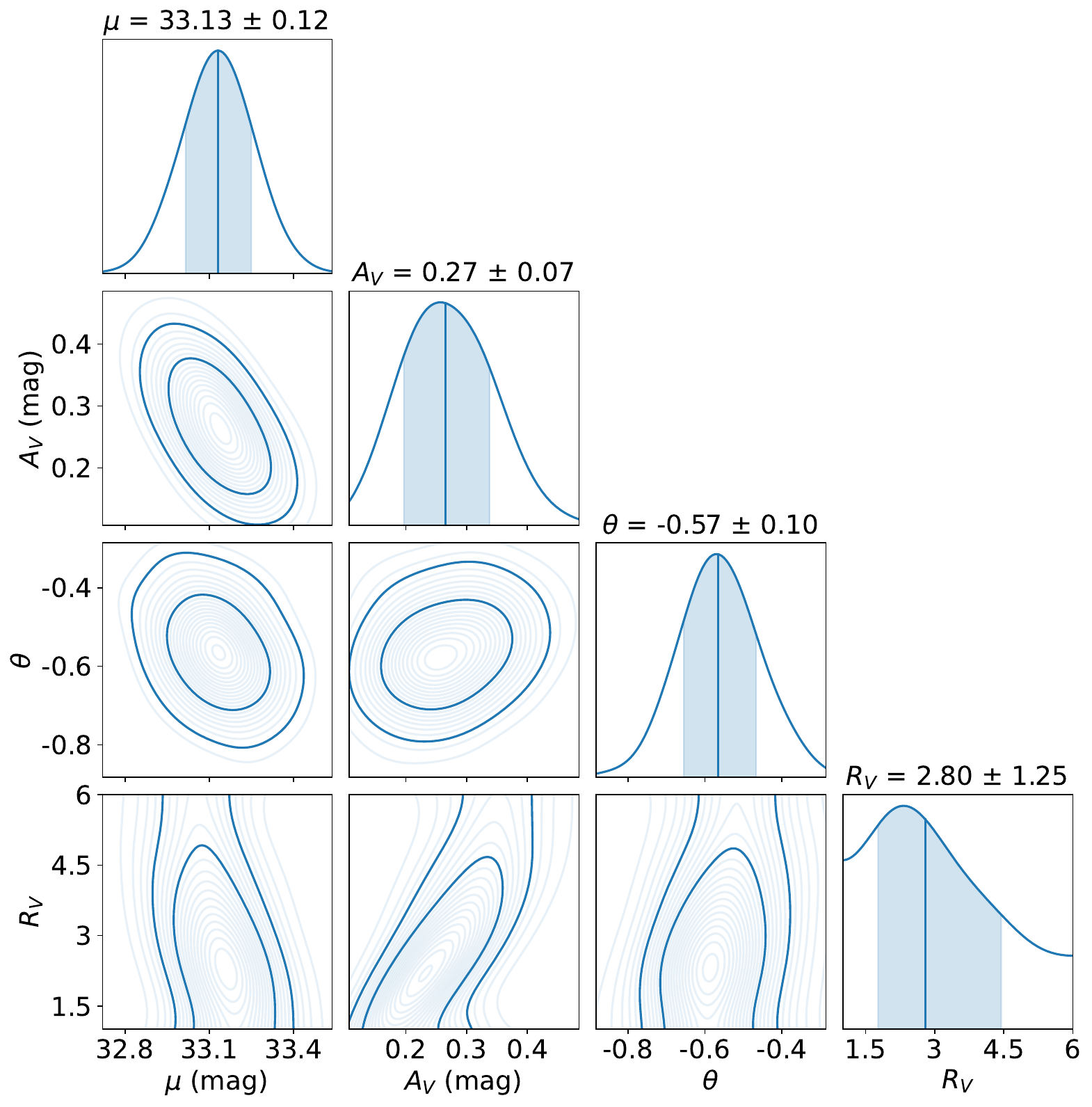}
    \caption{\textsc{BayeSN} fit to the SN~2021hpr $grizy$ photometry (Table~\ref{tab:21hprphotometry}), and kernel density estimates of the posterior distributions. 
    }
    \label{fig:21hprRVfit}
\end{figure}

We first fit each of NGC 3147's siblings individually with the new W22 \textsc{BayeSN} model. Fig.~\ref{fig:21hprRVfit} shows the SN~2021hpr fit posteriors of $(\mu, A_{V}, \theta, R_{V})$: the distance modulus, dust extinction, light curve shape, and dust law shape, respectively. Appendix~\ref{S:Appendix2021hprSpectrum} shows the SN~1997bq and SN~2008fv individual-fit posteriors.

\begin{figure}
    \centering
    \includegraphics[width=1\linewidth]{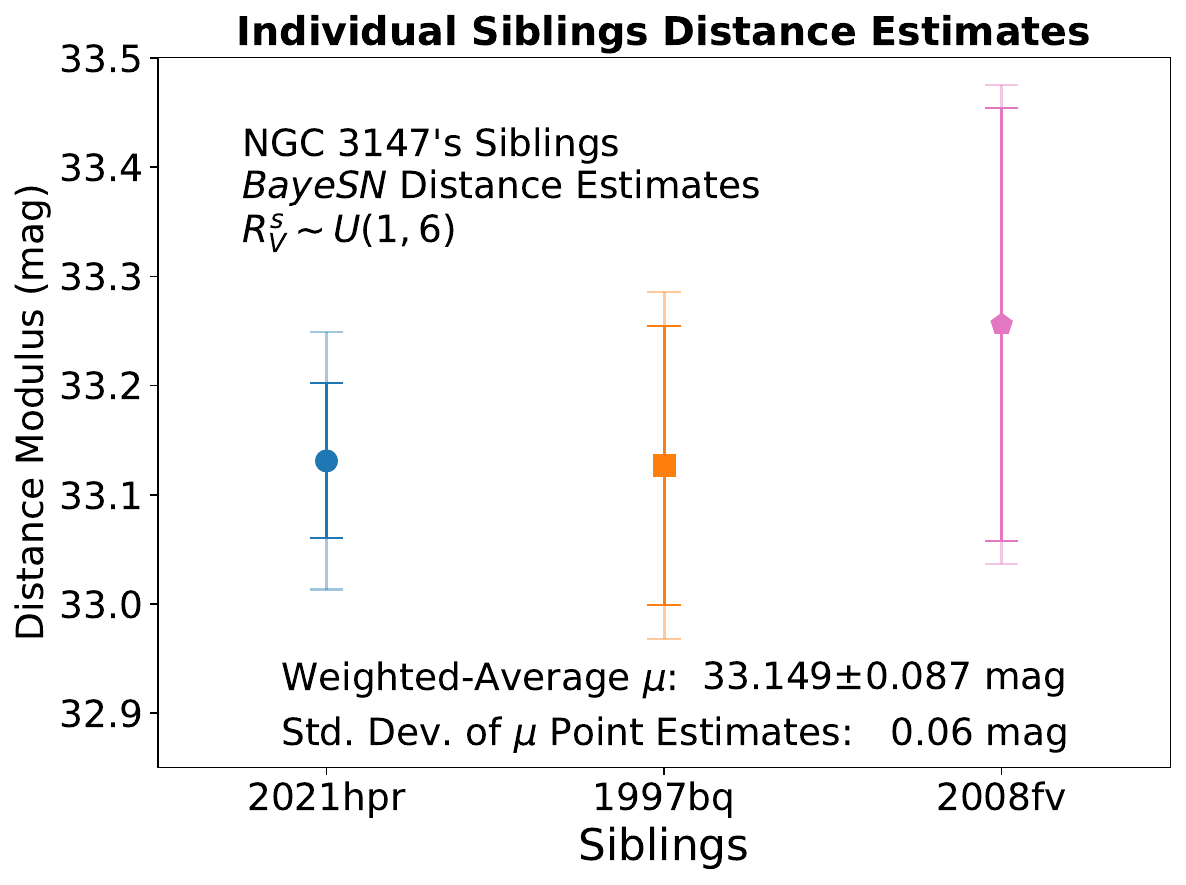}
    \includegraphics[width=1\linewidth]{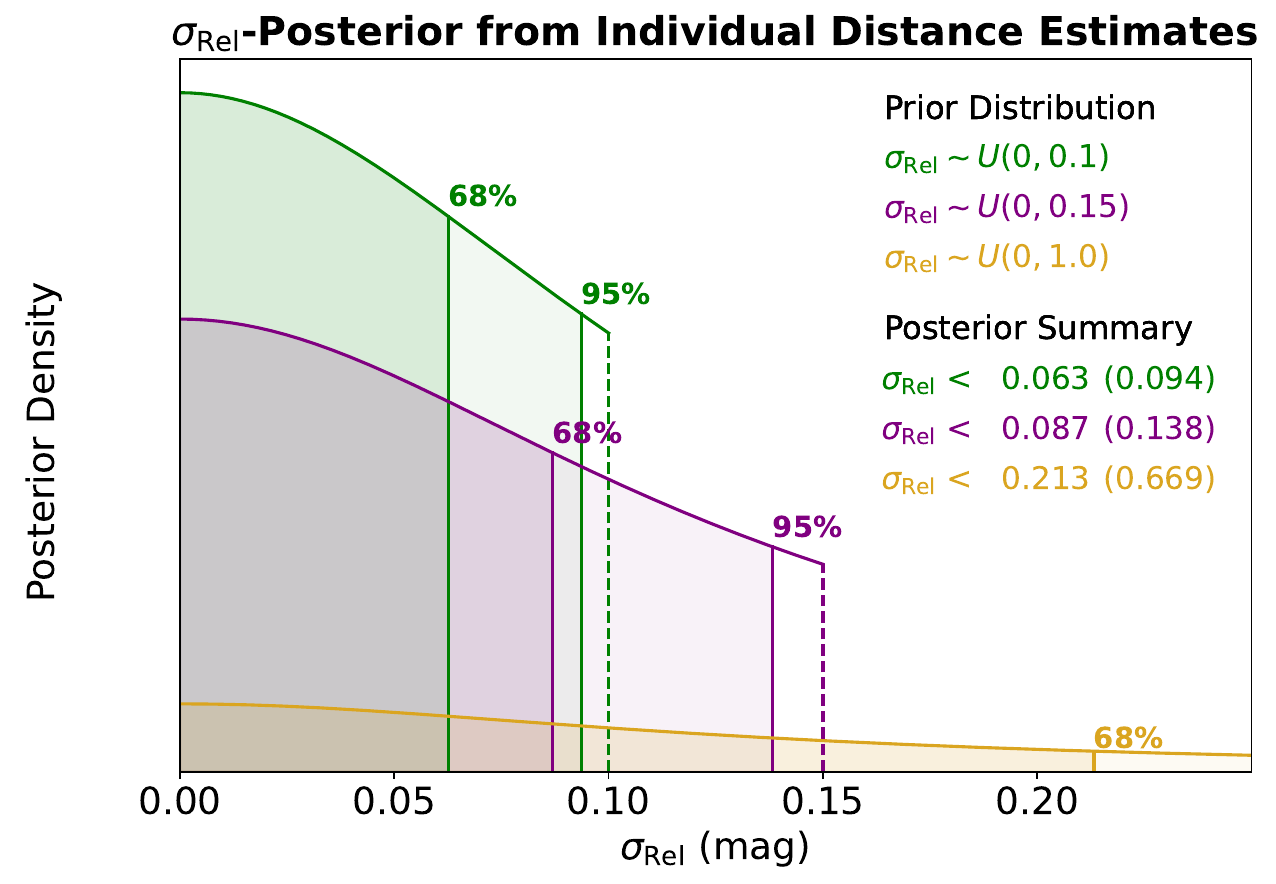}
    \includegraphics[width=1\linewidth]{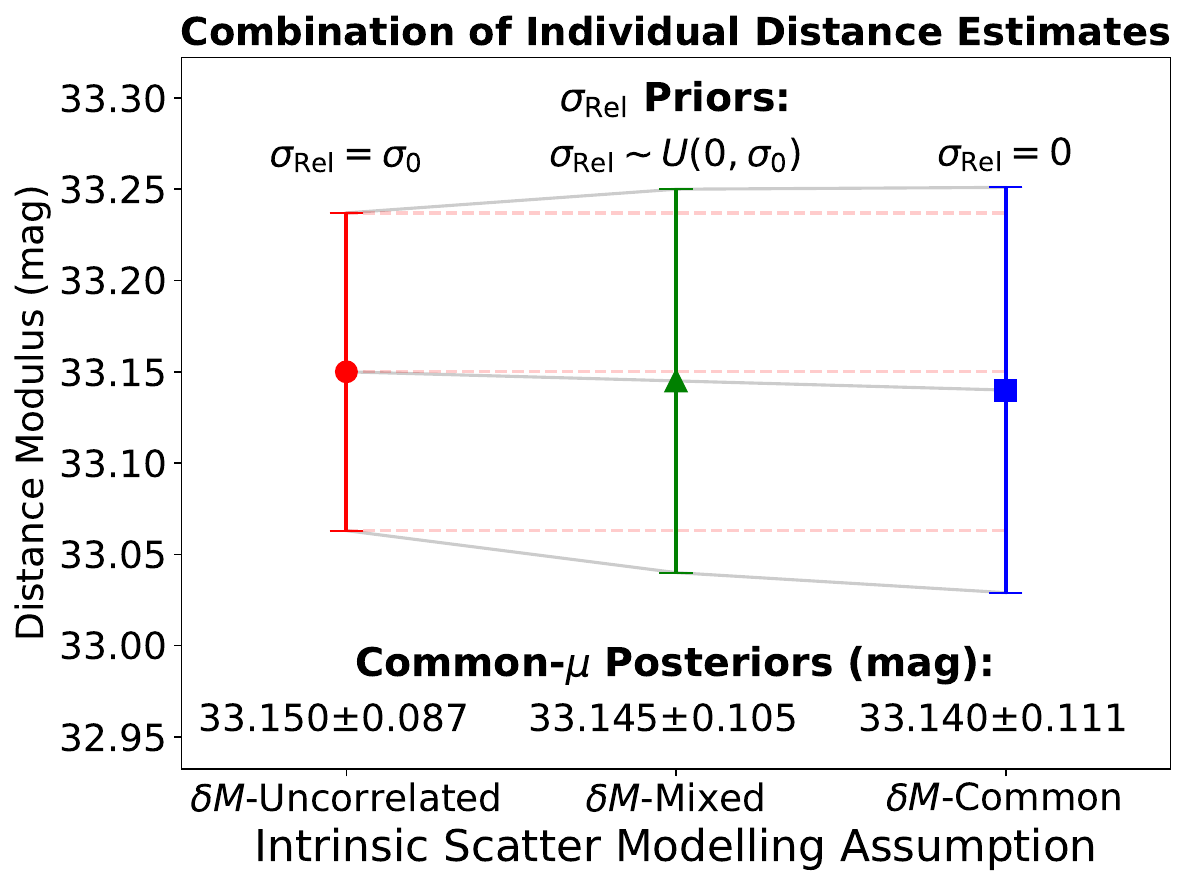}
    \caption{\textbf{Analysis of individual siblings distance estimates.}
   (top) \textsc{BayeSN} distance estimates from individual fits to the siblings trio; error bars are the distance measurement errors [bold], and the total errors that include $\sigma_0=0.094$~mag [faint]. (middle) Posteriors of $\sigma_{\rm{Rel}}$ computed using analytic formulae in \S\ref{S:ConstrainingSigmaRel}. Posteriors are weakly constraining, so more siblings galaxies are required to tightly constrain $\sigma_{\rm{Rel}}$. The 0.06~mag standard deviation is thus a poor estimator of $\sigma_{\rm{Rel}}$. (bottom) The individual distance estimates are combined under the three intrinsic scatter modeling assumptions (Table~\ref{tab:IntrinsicScatterModellingAssumptions}). The distance uncertainty is minimised when $\sigma_{\rm{Rel}}=\sigma_0$ is assumed, while the best estimate is obtained via $\sigma_{\rm{Rel}}$-marginalization. 
    }
    \label{fig:SigmaRelPosteriors}
\end{figure}

Table~\ref{tab:intialindepfits} shows posterior summaries. The distance estimates are strongly consistent with one another, with a 0.060~mag sample standard deviation of distance point estimates (Fig.~\ref{fig:SigmaRelPosteriors}). This consistency indicates the data are reliable and the model is robust. However, this does not directly evidence that $\sigma_{\rm{Rel}}<\sigma_0$.

Instead, the important information is captured by the posterior on $\sigma_{\rm{Rel}}$. We compute cosmology-independent $\sigma_{\rm{Rel}}$ posteriors using the individual distance estimates, and Eqs.~(\ref{eq:indepsigma0posterior1},~\ref{eq:indepsigma0posterior2}). We test the sensitivity to the choice of $\sigma_{\rm{Rel}}$ prior upper bound: $(0.1,0.15,1)$~mag. To estimate distance measurement errors, or \textsc{BayeSN} `fitting uncertainties', $\hat{\sigma}_{\rm{fit},s}$, we use
\begin{equation}
\label{eq:estimatesigmatlam}
    \hat{\sigma}_{\rm{fit},s} = \sqrt{\hat{\sigma}^2_{\mu,s}-\hat{\sigma}^2_{0,s}},
\end{equation}
where, $(\hat{\sigma}_{\mu,s}, \hat{\sigma}_{0,s})$ are the posterior standard deviations of $\mu_s$ and $\delta M_s$, respectively\footnote{The \textsc{BayeSN} fitting uncertainties are thus the contribution towards the individual distance uncertainties from fitting the time- and wavelength-dependent components.}. 

Fig.~\ref{fig:SigmaRelPosteriors} shows the $\sigma_{\rm{Rel}}$-posteriors. Although they peak at zero, the 68\%~(95\%) posterior upper bounds are strongly prior dependent, and extend out as far as $\sigma_{\rm{Rel}} < 0.21~(0.67)$~mag. 
This shows, unsurprisingly, that the $\sigma_{\rm{Rel}}$ constraints are weak, and the trio's data alone do not rule out that $\sigma_{\rm{Rel}}>\sigma_0$. Therefore, the sample standard deviation of distance point estimates is a poor indicator of $\sigma_{\rm{Rel}}$. More siblings galaxies are required to tightly constrain $\sigma_{\rm{Rel}}$. 

Nonetheless, we can robustly combine individual siblings distances by marginalizing over $\sigma_{\rm{Rel}}$, whilst imposing an informative prior, $\sigma_{\rm{Rel}}\sim U(0,\sigma_0)$. To perform this marginalization, we build three simple models in the probabilistic programming language \textsc{Stan}~\citep{Carpenter17, Stan20}, corresponding to the three $\delta M$ modeling assumptions in Table~\ref{tab:IntrinsicScatterModellingAssumptions}. We run each fit for 100,000 samples, which reduces the Monte Carlo error to $<1$~mmag. 

Results in Fig.~\ref{fig:SigmaRelPosteriors} show
the distance uncertainty is minimised, 0.087~mag, when adopting the $\delta M$-Uncorrelated assumption. Meanwhile, marginalizing over $\sigma_{\rm{Rel}}$ returns a larger and more robust distance uncertainty: 0.105~mag. This methodology is inexpensive, and can be implemented into cosmological analyses to improve joint siblings distance estimates.

\begin{figure*}[h]
    \centering
    \includegraphics[width=1\linewidth]{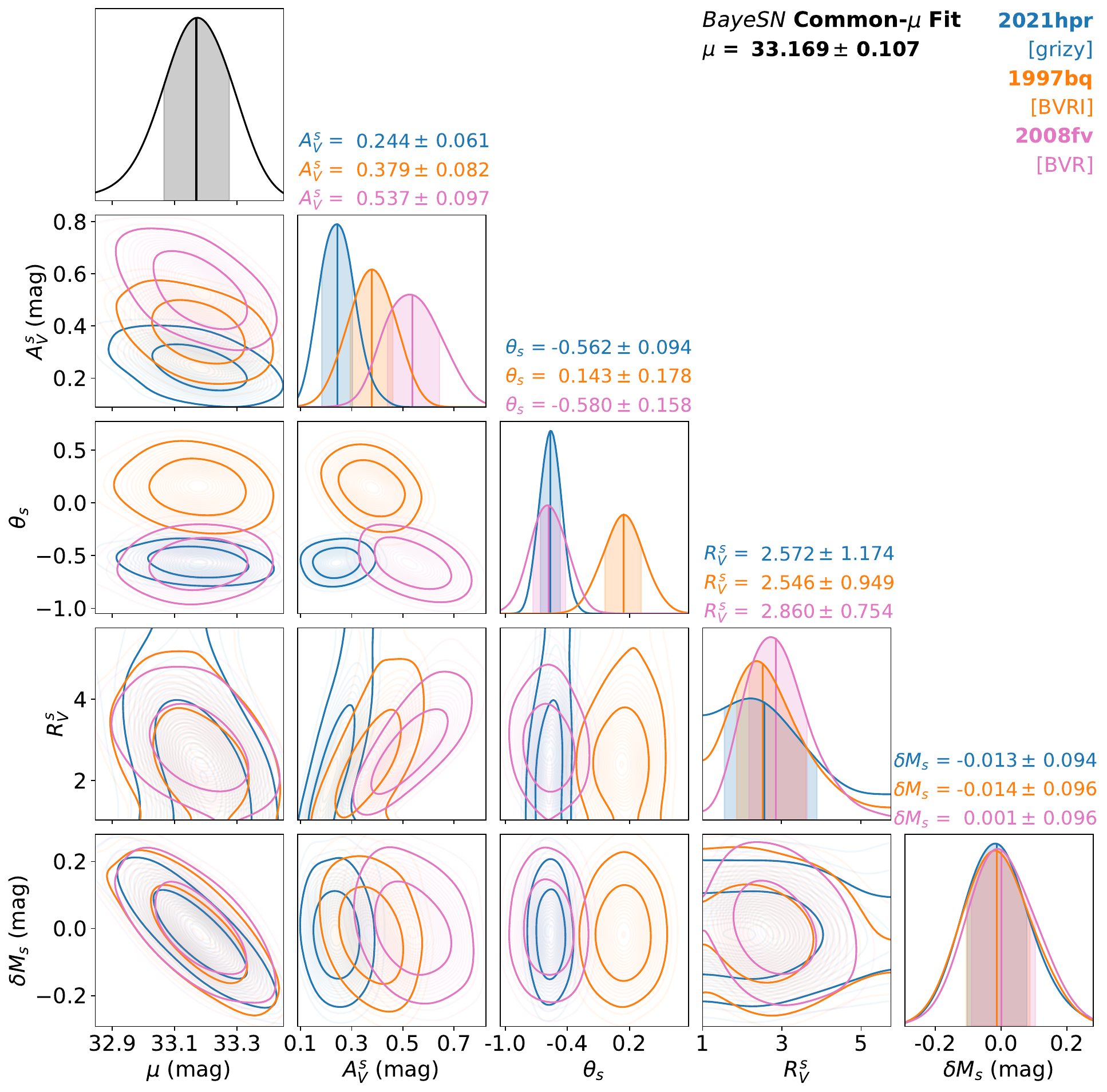}
    \caption{Posterior from the joint fit to NGC~3147's siblings' light curves. Constraints on the distance hyperparameter [black], and each sibling's individual parameters [colors], improve compared to individual fits. This fit marginalizes over $\sigma_{\rm{Rel}}$ with an informative prior, $\sigma_{\rm{Rel}}\sim U(0,\sigma_0)$, and $\sigma_0=0.094$~mag, to yield robust estimates of the distance and chromatic parameters. 
    }
    \label{fig:JointFit}
\end{figure*}

\begin{figure*}
    \centering
    \includegraphics[width=1\linewidth]{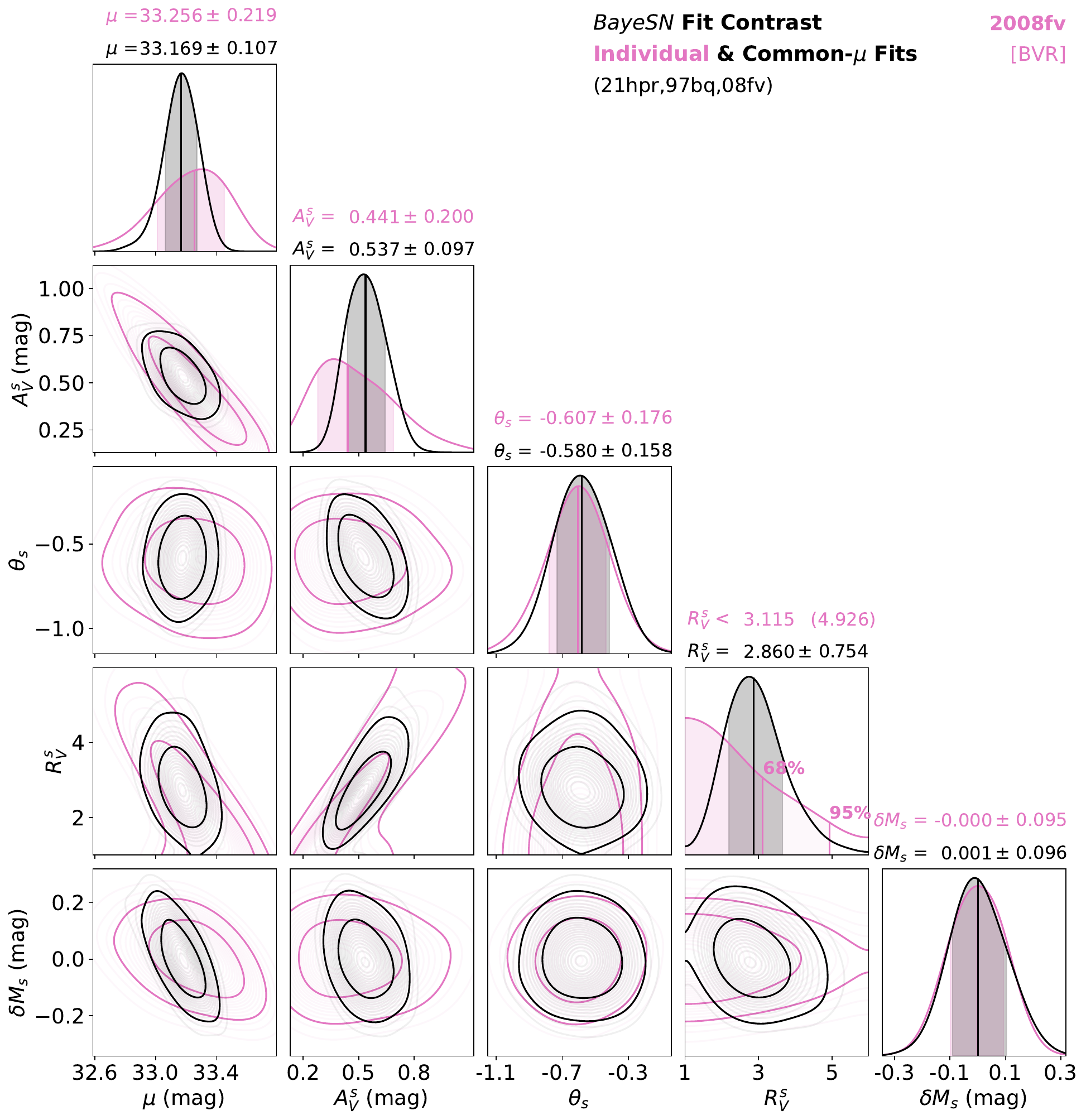}
    \caption{
    Contrast plot of SN~2008fv fit posteriors showcases the improvement in parameter estimates from the individual fit [pink] to the joint fit [black]. For example, the uncertainties on the individual dust parameters, $(A^s_V, R^s_V)$, shrink by $\approx (52,29)\%$. With only $BVR$ data in the individual fit, $R^s_V$ constraints are weak; however, the joint fit rules out low unphysical values $R^s_V\approx1$. 
    }
    \label{fig:ContrastPlot}
\end{figure*}

\subsection{\textsc{BayeSN} Joint Fits}
\label{S:JointFits}

Next, we apply the new \textsc{BayeSN} model architecture (\S\ref{S:ModellingAssumptions}), to jointly fit the siblings' light curves \textit{simultaneously}, while imposing that the siblings have a common, but unknown, distance. Fig.~\ref{fig:JointFit} displays the joint-fit posterior from marginalizing over $\sigma_{\rm{Rel}}$. As expected, the constraints on distance improve compared to individual fits (Table~\ref{tab:intialindepfits}). \textbf{Moreover, the constraints on each sibling's \textit{individual} parameters also improve by jointly fitting the siblings' light curves.}

We visualise this in Fig.~\ref{fig:ContrastPlot}, for SN~2008fv. This SN has optical $BVR$ data only, and so benefits greatly from the joint fit; in particular, the uncertainties in the individual dust parameters $(A^s_V, R^s_V)$ reduce by $\approx(52,29)\%$. Further, the individual-fit $R^s_V$ constraint peaks at the lower prior boundary $R^s_V=1$, while the joint fit rules out low and unphysical values, to constrain $R^s_V=2.86\pm0.75$.

The benefits of siblings data can be understood by drawing parallels with NIR data. Just as NIR data provide added leverage on the dust parameters, and hence the distance, so too siblings data provide added leverage on the distance, which improves constraints on the remaining parameters, like the dust parameters. 

\begin{figure}
    \includegraphics[width=1\linewidth]{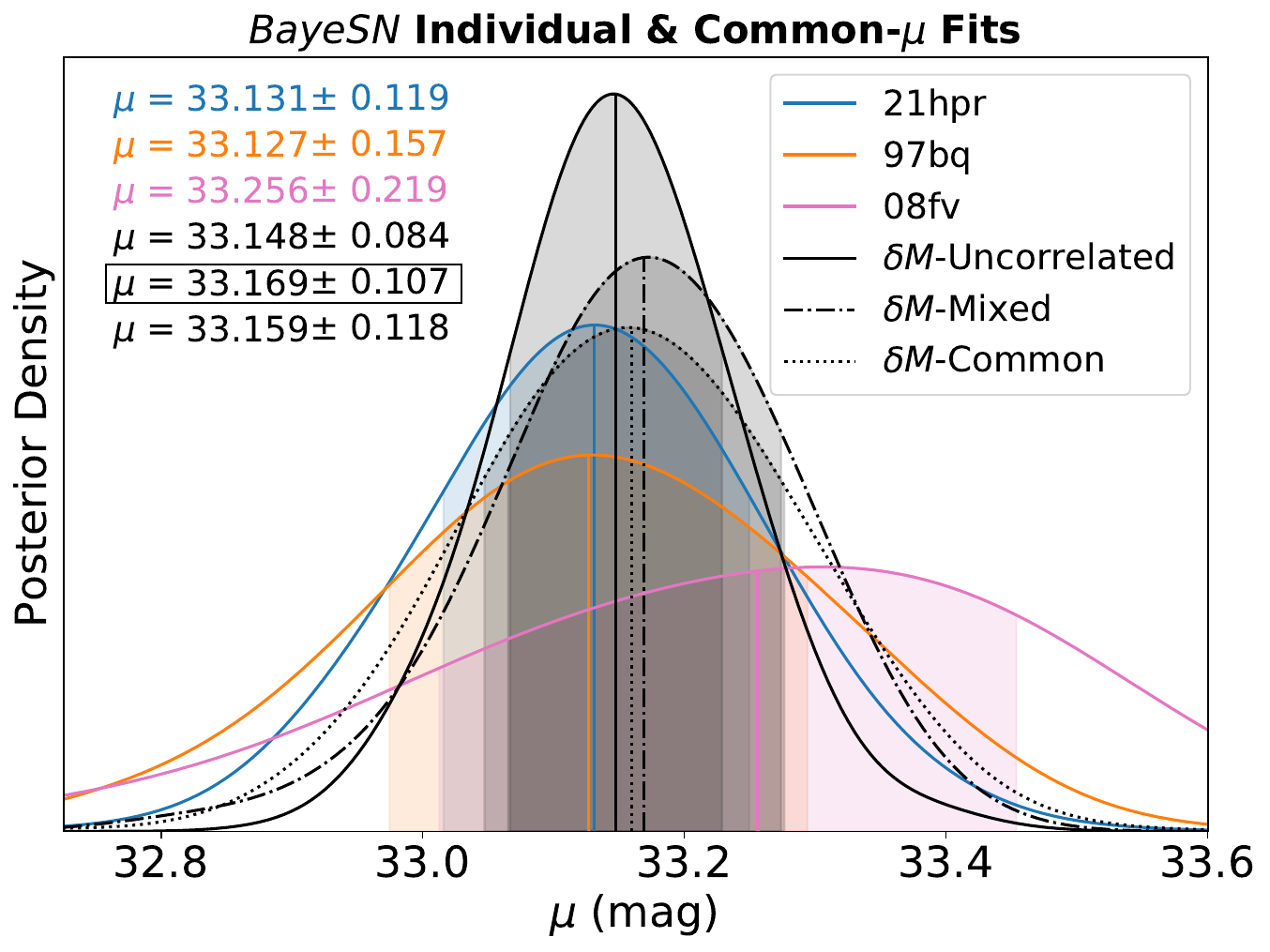}
    \includegraphics[width=0.9735\linewidth]{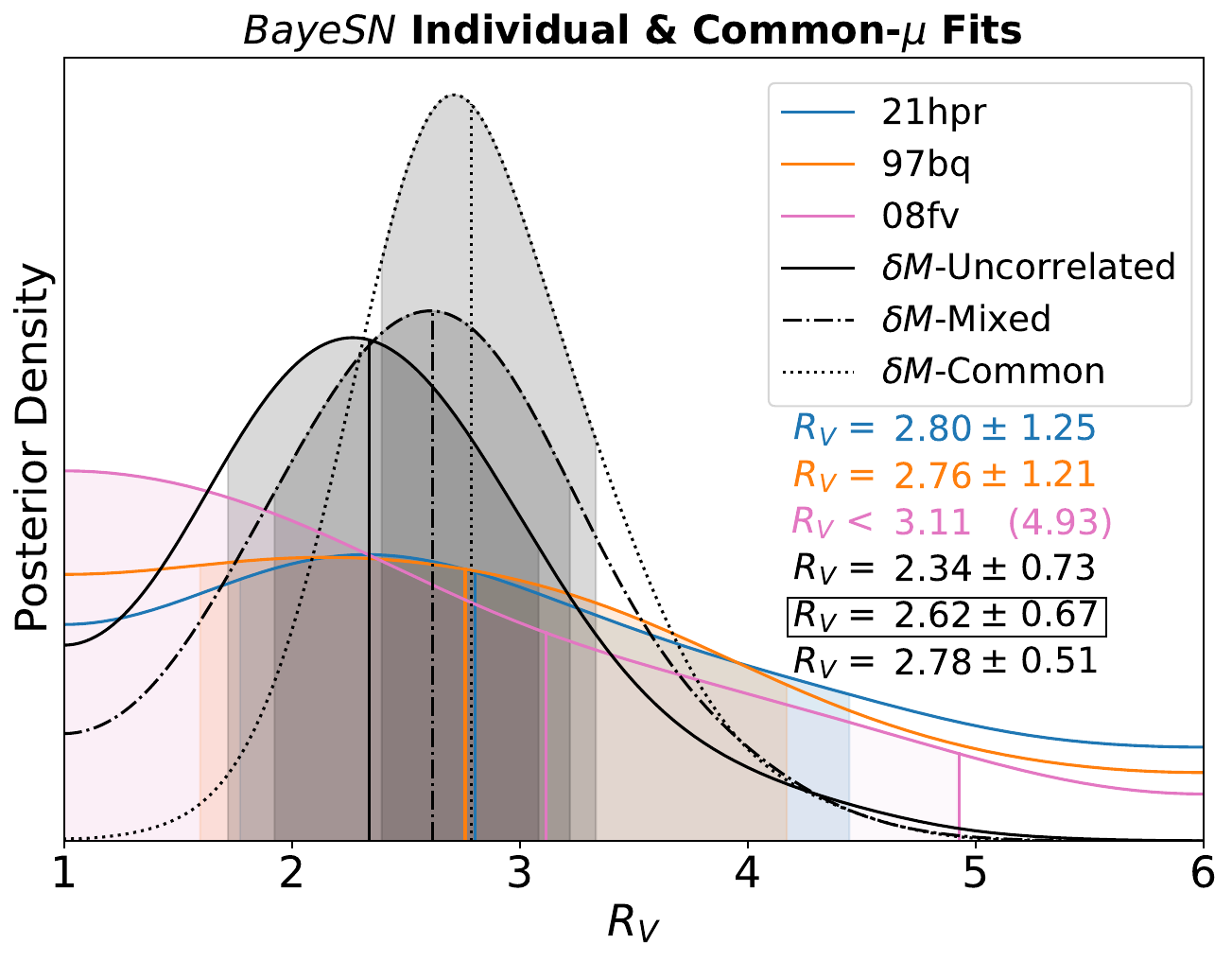}
    \caption{Chromatic parameter uncertainties are affected in the opposite sense to the distance hyperparameter by different intrinsic scatter modeling assumptions (Table~\ref{tab:IntrinsicScatterModellingAssumptions}). Overlay of distance and common-$R_V$ posteriors from individual fits [colors], and joint fits [black]. (top panel) Assuming the siblings are uncorrelated minimises the distance estimate uncertainty, while marginalizing over $\sigma_{\rm{Rel}}$ yields robust uncertainties. (bottom panel) The common-$R_V$ posteriors [black] show that the more correlated the intrinsic scatter components are, the less freedom there is in the fit, which results in tighter constraints on chromatic parameters. This is opposite to the behaviour of the distance hyperparameter. Marginalizing over $\sigma_{\rm{Rel}}$ yields $R_V=2.62\pm0.67$.
    }
    \label{fig:MuRVPosteriors}
\end{figure}

To more precisely constrain $R_V$ we assume it is the same for all three siblings. The common-$R_V$ constraint from marginalizing over $\sigma_{\rm{Rel}}$ is $R_V=2.62\pm0.67$. This is consistent with the W22 global value, $R_V = 2.659$, and the population mean, $R_V=2.70\pm0.25$, constrained in \cite{Thorp21}. Fig.~\ref{fig:MuRVPosteriors} shows the increase in precision on $R_V$ constraints, compared to individual fits. 

We further fit under the other two intrinsic scatter modeling assumptions: $\delta M$-Uncorrelated, and $\delta M$-Common. Fig.~\ref{fig:MuRVPosteriors} shows that the $R_V$ uncertainties behave in the opposite sense to the distance uncertainty: with larger $\sigma_{\rm{Rel}}$ values, there is a larger dispersion of $\delta M$ parameters, and more freedom in the fit, which leads to larger $R_V$ uncertainties. We check and assert that this trend also applies to all other chromatic parameters (like the dust extinction, $A_V^s$, and the light curve shape parameters, $\theta_s$). \textbf{Therefore, like the distance, the best chromatic parameter constraints are obtained by marginalizing over $\sigma_{\rm{Rel}}$.}

\subsection{$H_0$ Constraints}
\label{S:H0constraints}

Our work culminates in an estimate of the Hubble constant, obtained by simultaneously fitting the siblings trio's light curves, a Cepheid distance to NGC 3147~\citep{Riess22}, and \textsc{BayeSN} distance estimates to 109 Hubble flow SNe Ia. This inference is motivated by the conclusion in \cite{Scolnic20}, which states: ``Understanding the limited correlation between SNe in the same host galaxy will be important for SH0ES to properly propagate the combined uncertainty''. While understanding the root cause of correlated intrinsic scatter is beyond the scope of this work, our new relative scatter model marginalizes over this correlation, leading to robust uncertainties. Moreover, correlated intrinsic scatter was neglected in the recent siblings-$H_0$ estimate in \cite{Gallego-Cano22}. This sub-section thus demonstrates how siblings can be robustly modeled to estimate cosmological parameters.

We simultaneously fit the siblings' light curves with the Cepheid and Hubble flow distances, and we marginalize over $\sigma_{\rm{Rel}}$ using the informative prior: $\sigma_{\rm{Rel}} \sim U(0,\sigma_0).$ This extracts the most information out of the data via jointly fitting the siblings light curves, whilst accurately quantifying the distance uncertainty via $\sigma_{\rm{Rel}}$ marginalization. Appendix~\ref{S:AppendixH0Methods} details our $H_0$-inference methodologies further.

We estimate $H_0=78.4\pm 6.5\,\text{km\,s}^{-1}\text{\,Mpc}^{-1}$. The posterior is displayed in Fig.~\ref{fig:H0inference}. In Appendix~\ref{S:AppendixH0Methods}, we show this result is insensitive to the siblings' $R_V^s$ modeling assumptions, and their intrinsic scatter modeling assumptions. Assuming the siblings are uncorrelated yields $H_0=78.4 \pm 6.2\,\text{km\,s}^{-1}\text{\,Mpc}^{-1}$, meaning $\approx 1.7\,\text{km\,s}^{-1}\text{\,Mpc}^{-1}$ is added in quadrature to the $H_0$ uncertainty as result of marginalizing over $\sigma_{\rm{Rel}}$. Equivalently, this is a $\approx 2.2\%$ increase in the $H_0$ uncertainty.

Our Hubble constant estimate is consistent with typical local Universe measurements, in particular $H_0=73.04\pm1.04\,\text{km\,s}^{-1}\text{\,Mpc}^{-1}$~\citep{Riess22}, and also the lower \textit{Planck} value: $H_0=67.4\pm0.5\,\text{km\,s}^{-1}\text{\,Mpc}^{-1}$~\citep{Planck20}. The statistical error dominates, and including more calibrator galaxies will lead to a more accurate and precise $H_0$ inference, that is less sensitive to Cepheid modeling choices (as seen in figure 18 in \citealt{Riess22}; see companion paper: \citealt{Dhawan23}). The siblings' correlation is a sub-dominant effect in this case study; nonetheless, we have demonstrated, for the first time, how to robustly propagate the siblings' combined uncertainty to infer cosmological parameters.

\begin{figure*}
    \includegraphics[width=1\linewidth]{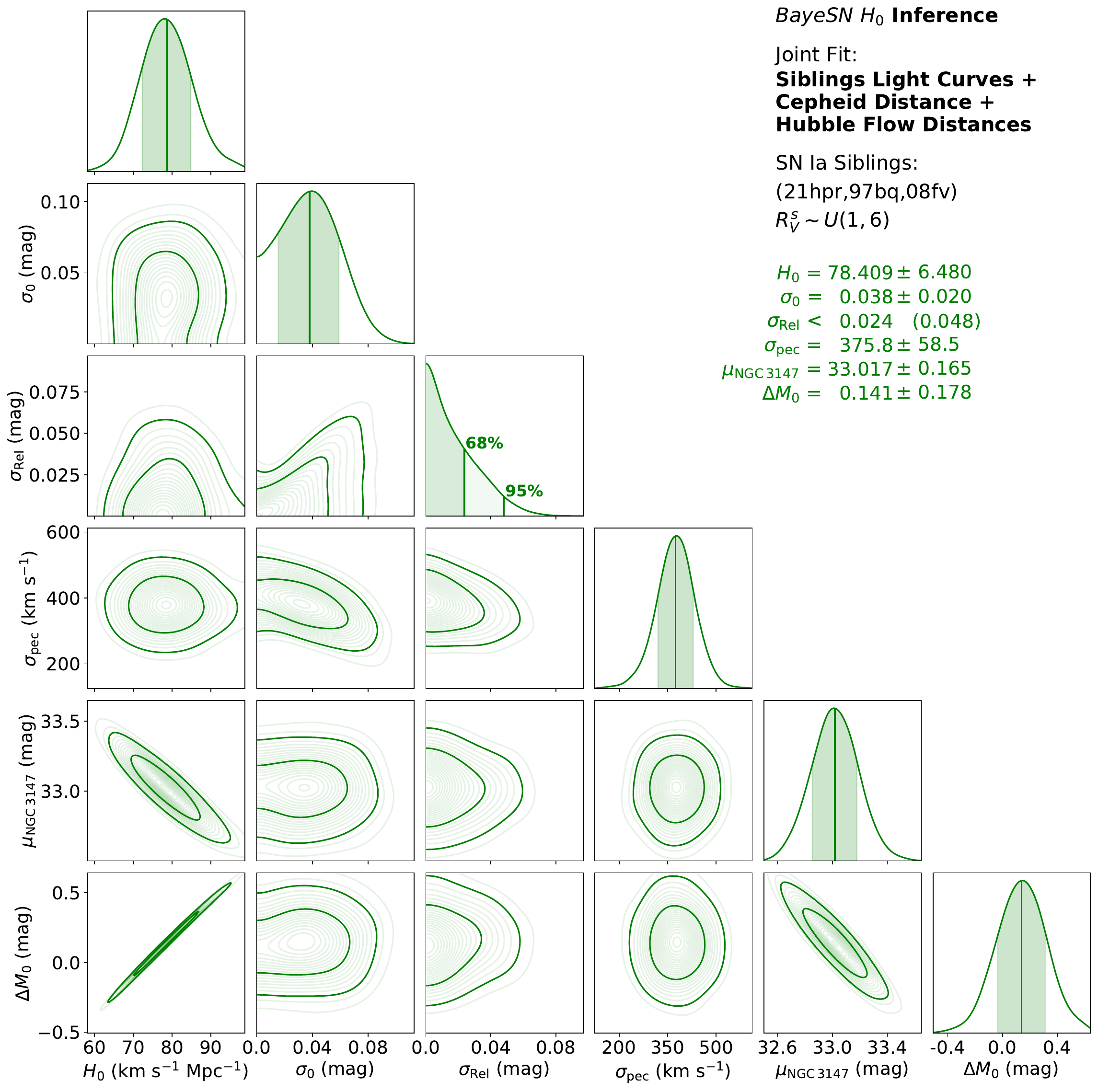}
    \caption{Hubble constant inference from simultaneously fitting NGC~3147's siblings' light curves, the Cepheid distance from \citet{Riess22}, and \textsc{BayeSN} distance estimates to 109 SNe Ia in the Hubble flow. We marginalize over the total intrinsic scatter, $\sigma_0$, the siblings' relative intrinsic scatter, $\sigma_{\rm{Rel}}$, and the peculiar velocity dispersion, $\sigma_{\rm{pec}}$. Marginalizing over $\sigma_{\rm{Rel}}$ while jointly fitting the siblings' light curves yields the best NGC~3147 distance constraint. We estimate $H_0=78.4\pm 6.5\,\text{km\,s}^{-1}\text{\,Mpc}^{-1}$.}
    \label{fig:H0inference}
\end{figure*}

\vspace{-0.1cm}
\section{Conclusions}
\label{S:Conclusions}

We have analyzed light curves of the SN~Ia trio of siblings in the high-stellar-mass Cepheid calibrator galaxy NGC~3147. This includes PS1 $grizy$ photometry of SN~2021hpr from the Young Supernova Experiment~\citep{Jones21}, presented in Table~\ref{tab:21hprphotometry}. We use a new publicly available W22 version of the \textsc{BayeSN} hierarchical optical-to-near-infrared SN~Ia SED model (\S\ref{S:NewModel}, Appendix~\ref{AppendixNewModel}),
which we have retrained simultaneously on $BgVrizYJH$ (0.35--1.8\,$\mu$m) data of 236 SNe~Ia. 

We summarise the key conclusions from this work:
\begin{itemize}

    \item The relative intrinsic scatter, $\sigma_{\rm{Rel}}$, is the intrinsic scatter of individual siblings distance estimates relative to another within a galaxy. It quantifies the contribution towards the total intrinsic scatter, $\sigma_0$, from within-galaxy variations about the siblings' common properties, and is constrained without assuming any cosmology. Therefore, it is distinct from $\sigma_0$, and we expect $\sigma_{\rm{Rel}}\leq\sigma_0$.

     \item We present analytic formulae for computing a $\sigma_{\rm{Rel}}$ posterior from individual siblings distance estimates (\S\ref{S:ConstrainingSigmaRel}). Applying this to NGC 3147's siblings, the posterior is wide, meaning the sample standard deviation of distance point estimates is a poor indicator of $\sigma_{\rm{Rel}}$, particularly in this small sample size limit.
    
    \item Assuming $\sigma_{\rm{Rel}}=\sigma_0$ will underestimate the uncertainty on a joint siblings distance estimate if in fact $\sigma_{\rm{Rel}}<\sigma_0$. An inexpensive way of constraining $\mu$ from individual siblings distance estimates is to marginalize over $\sigma_{\rm{Rel}}$ with an informative prior: $\sigma_{\rm{Rel}}\sim U(0,\sigma_0)$. 
    
    \item For the first time, we hierarchically fit siblings light curves simultaneously, to estimate a common, but unknown, distance hyperparameter. The \textsc{BayeSN} joint fit benefits from a hierarchical sharing of information, returning improved constraints on the common distance, as well as each sibling's \textit{individual} chromatic parameters (e.g. light curve shape, host galaxy dust parameters). For example, SN~2008fv's individual-fit $R_V^s$ constraint peaks at $R^s_V=1$, but the joint fit naturally rules out low unphysical values to constrain $R^s_V=2.86\pm0.75$.

    \item Chromatic parameters are also affected by $\sigma_{\rm{Rel}}$, and in the opposite sense to the distance, with larger $\sigma_{\rm{Rel}}$ values returning \textit{larger} chromatic parameter uncertainties. Therefore, $\sigma_{\rm{Rel}}$-marginalization yields robust estimates of chromatic parameters, as well as the distance. We infer a common dust hyperparameter $R_V=2.62\pm0.67$.
    
    \item We estimate the Hubble constant, $H_0=78.4\pm 6.5\,\text{km\,s}^{-1}\text{\,Mpc}^{-1}$, by hierarchically fitting the siblings light curves with a Cepheid distance and Hubble flow distances, whilst marginalizing over $(\sigma_{\rm{Rel}},A_V^s,R_V^s)$ in the calibrator galaxy, and $(\sigma_0,\sigma_{\rm{pec}})$ in the Hubble flow. Marginalizing over $\sigma_{\rm{Rel}}$ adds $\approx 1.7\,\text{km\,s}^{-1}\text{\,Mpc}^{-1}$, or $\approx 2.2\%$, in quadrature to the $H_0$ uncertainty, compared to assuming $\sigma_{\rm{Rel}}=\sigma_0$. This inference is the first cosmological analysis to robustly propagate the combined uncertainty from SN~Ia siblings.
\end{itemize}

This work provides a robust and principled Bayesian framework for hierarchically analyzing SN~Ia siblings. Consequently, we have shown the relative intrinsic scatter, $\sigma_{\rm{Rel}}$, is a key parameter in any siblings analysis. Marginalizing over $\sigma_{\rm{Rel}}$ yields robust inferences of siblings distances for cosmology, and chromatic parameters for SN-host correlation studies. With more siblings galaxies, there is the potential to tightly constrain $\sigma_{\rm{Rel}}$, and compare it against $\sigma_0$, to investigate the dominant systematics acting in \textsc{BayeSN} and other SN~Ia models. 

A multi-galaxy $\sigma_{\rm{Rel}}$ inference may be sensitive to cross-calibration systematics originating from siblings observed on different photometric systems. While extensive work has gone in to building distance covariance matrices for SALT2 distances in the Pantheon+ sample~\citep{Scolnic21,Brout22}, building analogous covariance matrices for \textsc{BayeSN} is beyond the scope of this work. A workaround will be to analyze a sample of SN~Ia siblings observed on a single photometric system, thus mitigating cross-calibration systematics. The near-future sample of ZTF DR2 siblings~\citep[][]{Dhawan21, Graham22} is expected to contain $\approx 30$ SN~Ia siblings galaxies, so will be an interesting test set for investigating $\sigma_{\rm{Rel}}$, and its SN-model dependence.

Beyond constraining $\sigma_{\rm{Rel}}$, siblings are useful for studying within-galaxy dispersions of other features, like $(A_V^s, \theta_s, R_V^s)$, and host properties~\citep{Scolnic20, Scolnic21, Biswas22}. Identifying and studying features that have within-galaxy dispersions which are smaller than their total dispersions may help to explain any correlations amongst siblings distance estimates.

Intermediate sized samples of SN~Ia siblings have already been compiled, numbering $\sim 10-40$ siblings~(e.g. \citealp{Burns20, Scolnic20, Scolnic21, Graham22}). The largest siblings sample to date was recently presented in \cite{Kelsey23}, and comprises 158 galaxies with 327 SNe Ia. The potential for discovering more siblings is promising, with surveys such as the Dark Energy Survey (DES; \citealp{Scolnic20, Smith20}), the Legacy Survey of Space and Time (LSST;~\citealp{LSST18,Ivezic19,Scolnic20}), the Nancy Grace Roman Space
Telescope (\textit{Roman}; \citealp{Hounsell18, Rose21}), the Young Supernova Experiment (YSE; \citealp{Jones21}) and the Zwicky Transient Facility (ZTF; \citealp{Dhawan21, Graham22}). Therefore, SN~Ia siblings will be increasingly valuable for studying SN~Ia standardisation systematics in cosmological analyses in the years ahead.  


\section{Acknowledgments}.
\begin{acknowledgments}
We thank Adam G. Riess for providing the Cepheid distance to NGC~3147 prior to its public release on 18$^{\rm{th}}$ July 2022. 

S.M.W. was supported by the UK Science and Technology Facilities Council (STFC). S.T. was supported by the Cambridge Centre for Doctoral Training in Data-Intensive Science funded by STFC. K.S.M. acknowledges funding from the European Research Council under the European Union’s Horizon 2020 research and innovation programme (ERC Grant Agreement No. 101002652). This project has also been made possible through the ASTROSTAT-II collaboration, enabled by the Horizon 2020, EU Grant Agreement No. 873089. S.D. acknowledges support from the Marie Curie Individual Fellowship under grant ID 890695 and a Junior Research Fellowship at Lucy Cavendish College. D.O.J. acknowledges support by NASA through the NASA Hubble Fellowship grant HF2-51462.001 awarded by the Space Telescope Science Institute, which is operated by the Association of Universities for Research in Astronomy, Inc., for NASA, under contract NAS5-26555. The UCSC team is supported in part by NASA grant NNG17PX03C, NSF grant AST--1815935, the Gordon \& Betty Moore Foundation, the Heising-Simons Foundation, and by a fellowship from the David and Lucile Packard Foundation to R.J.F. G.N. was supported by the University of Illinois at Urbana-Champaign and the Center for Astrophysical Surveys at the National Center for Supercomputing Applications. Pan-STARRS is a project of the Institute for Astronomy of the University of Hawaii, and is supported by the NASA SSO Near Earth Observation Program under grants 80NSSC18K0971, NNX14AM74G, NNX12AR65G, NNX13AQ47G, NNX08AR22G, and by the State of Hawaii. K.D. acknowledges support in part by the NSF through grant AST-2108676.
N.E. acknowledges funding from the Center for Astrophysical Surveys Fellowship through the National Center for Supercomputing Applications at the University of Illinois at Urbana-Champaign.
A.G. is supported by the National Science Foundation Graduate Research Fellowship Program under Grant No. DGE–1746047. A.G. further acknowledges funding from the Center for Astrophysical Surveys Fellowship at UIUC/NCSA and the Illinois Distinguished Fellowship. This work was supported by a VILLUM FONDEN Investigator grant (project number 16599 and 25501). This work made use of the Illinois Campus Cluster, a computing resource that is operated by the Illinois Campus Cluster Program (ICCP) in conjunction with the National Center for Supercomputing Applications (NCSA) and which is supported by funds from the University of Illinois at Urbana-Champaign.
\end{acknowledgments}

%

\vspace{5mm}
\facilities{Pan-STARRS-1, HST}





\appendix
\section{W22 \textsc{BayeSN} Model}
\label{AppendixNewModel}
\begin{figure}
    \includegraphics[width=0.5\linewidth]{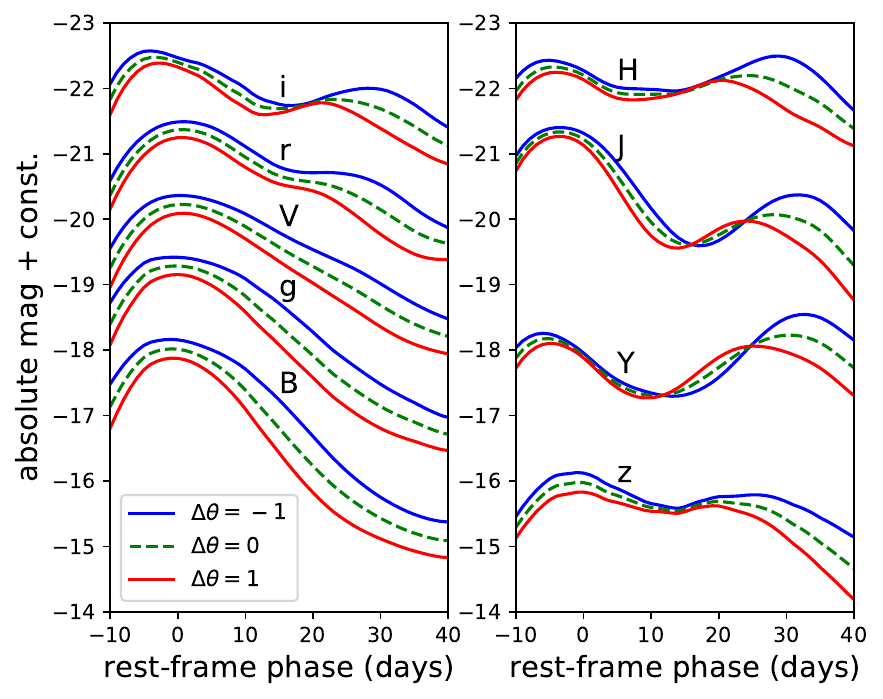}
    \includegraphics[width=0.5\linewidth]{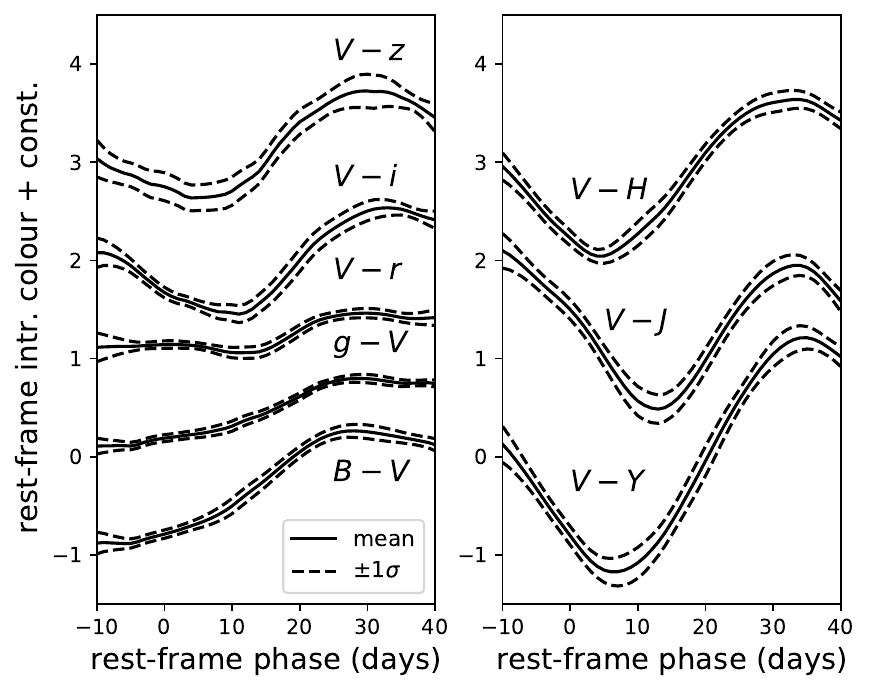}
    \caption{(left panel) Rest-frame light curves in $BgVrizYJH$ passbands are synthesised using the new W22 \textsc{BayeSN} model, to illustrate the intrinsic effect of $W_1(t,\lambda)$, the first functional principal component (FPC). Dust extinction and residual functions are set to zero, and the coefficient on the first FPC, $\theta$, is varied about the population mean; this variation gives rise to the \citet{Phillips93} `broader-brighter' empirical relation in optical passbands, while also affecting the NIR amplitudes and times of the second peak. For visual clarity, the $BgVrizYJH$ light curves have been vertically offset by (1.25, 0, -1, -2.25, -4, 2.5, 0.25, -2.75, -4)~mag, respectively.
    (right panel) Rest-frame intrinsic color curves are synthesised using the new W22 \textsc{BayeSN} model, to  illustrate the predicted level of residual intrinsic color variation after correction for dust extinction and light curve shape. Dust extinction and the light curve shape, $\theta$, are set to zero. The solid black lines depict the mean intrinsic color curves, while the dashed lines show a 1~$\sigma$ variation about the mean, computed by taking the standard deviation of many residual function realisations, each drawn from the intrinsic SED residual covariance matrix. For visual clarity, the colors are offset by (-0.75, 0.25, 1.25, 2.5, 3.5, 0.25, 2.25, 3.25)~mag for ($B-V$, $g-V$, $V-r$, $V-i$, $V-z$, $V-Y$, $V-J$, $V-H$), respectively.}
    \label{fig:W22FPCs}
\end{figure}

Here we describe the training and performance of the new W22 version of \textsc{BayeSN}\footnote{\url{https://github.com/bayesn/bayesn-model-files}}. The W22 training sample combines the \citetalias{Thorp21} and \citetalias{Mandel22} training samples. The \citetalias{Mandel22} training sample comprises 79 SNe Ia from \cite{Avelino19}, observed in $BVRIYJH$ passbands, compiled across a range of telescopes and surveys~\citep{Jha99, Krisciunas03,Krisciunas04:4sn,Krisciunas04:7sn,Krisciunas07, Stanishev07:03du,Pignata08:02dj,Wood-Vasey08, Hicken09:lc, Hicken12, Leloudas09, Friedman15, Krisciunas17}. This sample is approximately 80\% high-mass ($\log_{10}(\rm{M}_{*}/\rm{M}_{\odot}) > 10$). The \citetalias{Thorp21} training sample comprises 157 SNe~Ia from the untargeted Foundation DR1 survey, observed in $griz$ passbands with the Pan-STARRS-1 system~\citep{Foley18:found, Jones19}. This sample has a 48:109 split at $\log_{10}(\rm{M}_{*}/\rm{M}_{\odot}) \approx 10$, and a median mass at $\log_{10}(\rm{M}_{*}/\rm{M}_{\odot}) \approx 10.331$.

The training procedure, and all hyperpriors, are exactly as described in \cite{Mandel22}, with the exception of the priors on the intrinsic residual standard deviation vector, $\bm{\sigma}_\epsilon$. Where \cite{Mandel22} used an uninformative half-Cauchy prior with unit scale on each element of this vector, here, we place an uninformative half-normal prior with a scale of 0.15, i.e. $P(\sigma_{\epsilon,q})=\text{Half-}\mathcal{N}(\sigma_{\epsilon,q}|\mu=0,\sigma=0.15)$ for the $q$th element. This provides better regularization in the $z$-band and NIR, and can be interpreted as assuming that the residual scatter in any band at any given time is $<0.3$~mag with 95\% prior probability.

We verify the robustness of the W22 model by computing the $\rm{RMS}$ and $\sigma_{\rm{-pv}}$ values in the Hubble diagram, from fits to the full sample of 236 SNe~Ia\footnote{The $\sigma_{\rm{-pv}}$ statistic is an estimate of the dispersion in the Hubble residuals removing the expected contribution from peculiar velocity uncertainties (defined in equation 32; \citealp{Mandel22}, and adopting  $\sigma_{\rm{pec}}=150\,\text{km\,s}^{-1}$).}. The ($\rm{RMS}, \sigma_{\rm{-pv}}$) values are (0.129, 0.113)~mag. From W22 fits to each of the \citetalias{Thorp21} and \citetalias{Mandel22} samples, the values are (0.129, 0.117)~mag and (0.130, 0.104)~mag, respectively. This compares well to the Hubble diagram statistics obtained from fitting \citetalias{Thorp21}, (0.124, 0.112)~mag, and \citetalias{Mandel22}, (0.137, 0.109)~mag, to their respective training samples. When fitting W22 to the 40 SN Ia `NIR@max' sub-sample from \cite{Mandel22}, the fit statistics are (0.093, 0.085)~mag, as compared to the \citetalias{Mandel22} values: (0.096, 0.083)~mag.

The effect of the first functional principal component (FPC) is showcased in the left panels of Fig.~\ref{fig:W22FPCs}, and the population variations in intrinsic color are showcased in the right panels of Fig.~\ref{fig:W22FPCs}. The posterior means of the population hyperparameters learned in W22 model training (in addition to the FPCs and residual covariance matrix) are: the global dust law shape, $R_V = 2.659$, the dust extinction hyperparameter, $\tau_A = 0.252$~mag, and the total intrinsic scatter, $\sigma_0=0.094$~mag.

\section{Additional Siblings Data \& Fits}
\label{S:Appendix2021hprSpectrum}

The SN~2021hpr spectrum was obtained with the Kast spectrograph on the Lick Shane telescope \citep{KAST} on 13$^{\rm{th}}$ April 2021, $\approx 4.2$ days before $B$-band maximum brightness. The data were reduced with standard CCD processing and extractions using a custom data reduction pipeline\footnote{\url{https://github.com/msiebert1/UCSC\_spectral\_pipeline}} which employs IRAF\footnote{IRAF was distributed by the National Optical Astronomy Observatory, which was managed by the Association of Universities for Research in Astronomy (AURA) under a cooperative agreement with the National Science Foundation.}. We fit low-order polynomials to calibration-lamp spectra and perform small shifts based on sky lines in object frames to determine a wavelength solution. We employ custom Python routines and spectrophotometric standard stars to flux calibrate the spectrum and remove telluric lines \citep{Silverman12:bsnip}. We remove data and linearly interpolate the spectrum between 5650~\AA\ and 5710~\AA\ in the observer frame so as to remove a ghosting artifact. 

Fig.~\ref{fig:97bq08fvRVfit} shows individual \textsc{BayeSN} fits to the Pantheon+~\citep[][]{Scolnic21} light curves of SNe~1997bq and 2008fv. Table~\ref{tab:RVfixedfits} records posterior summaries from fixed-$R_V$ individual fits to the trio ($R_V^*=2.659$).

\begin{figure}
  \begin{minipage}[c]{0.5\textwidth}
    \includegraphics[width=\textwidth]{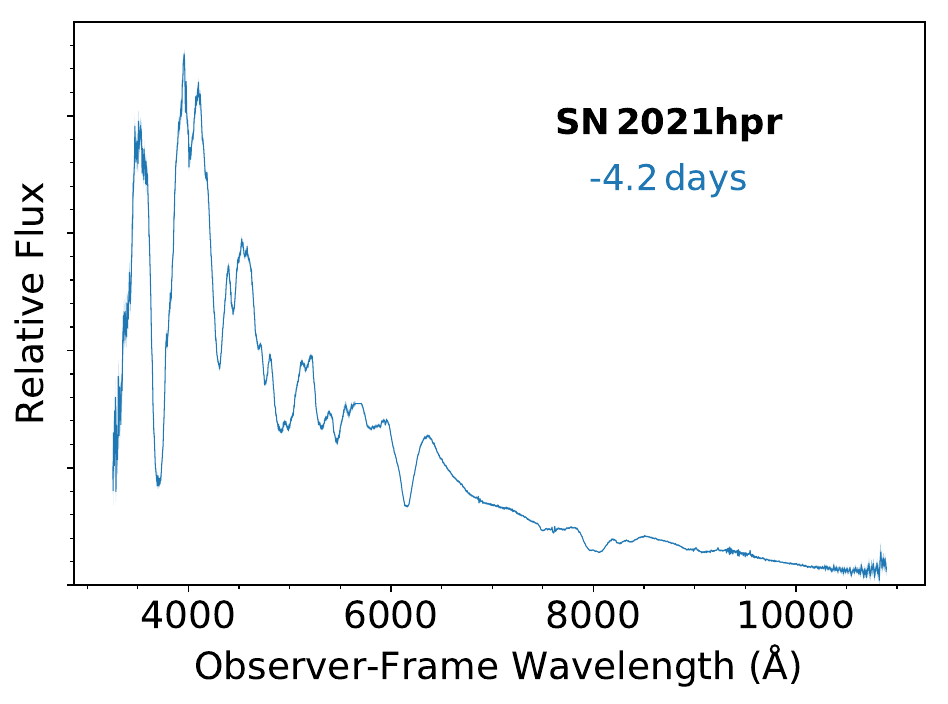}
  \end{minipage}\hfill
  \begin{minipage}[c]{0.5\textwidth}
    \caption{
       Spectrum of SN~2021hpr, observed $\approx 4.2$~d before the $B$-band maximum, using the Kast spectrograph on the Lick Shane telescope, in the observer-frame wavelength range: 3254--10894~\AA.
    } \label{fig:21hprspectrum}
  \end{minipage}
\end{figure}

\begin{figure*}
\gridline{\fig{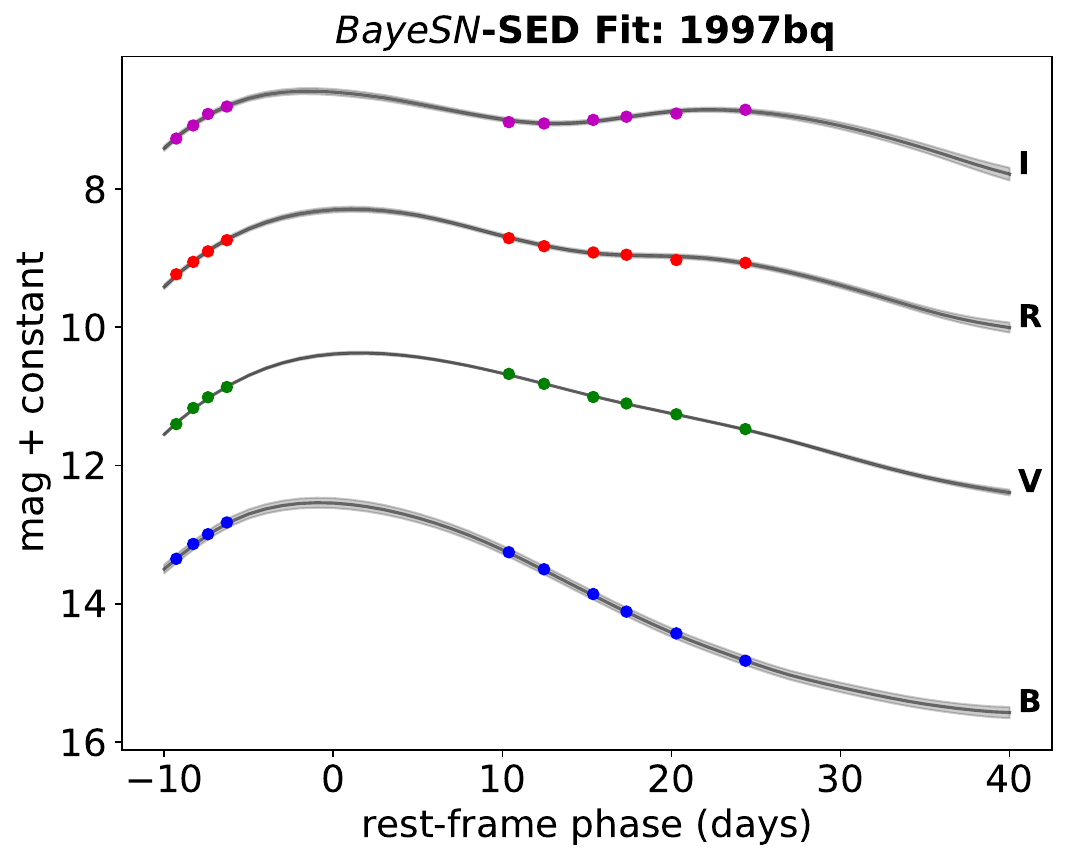}{0.5\textwidth}{}
          \fig{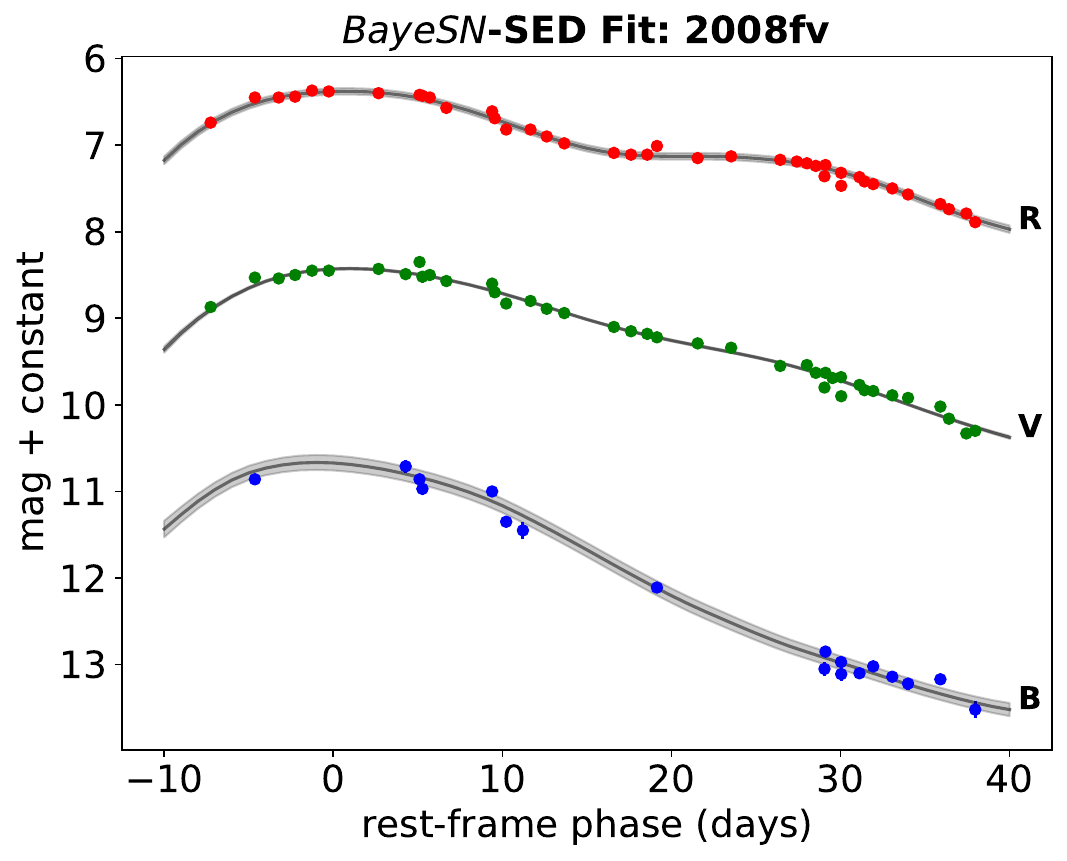}{0.5\textwidth}{}}
\vspace{-1cm}
\gridline{\fig{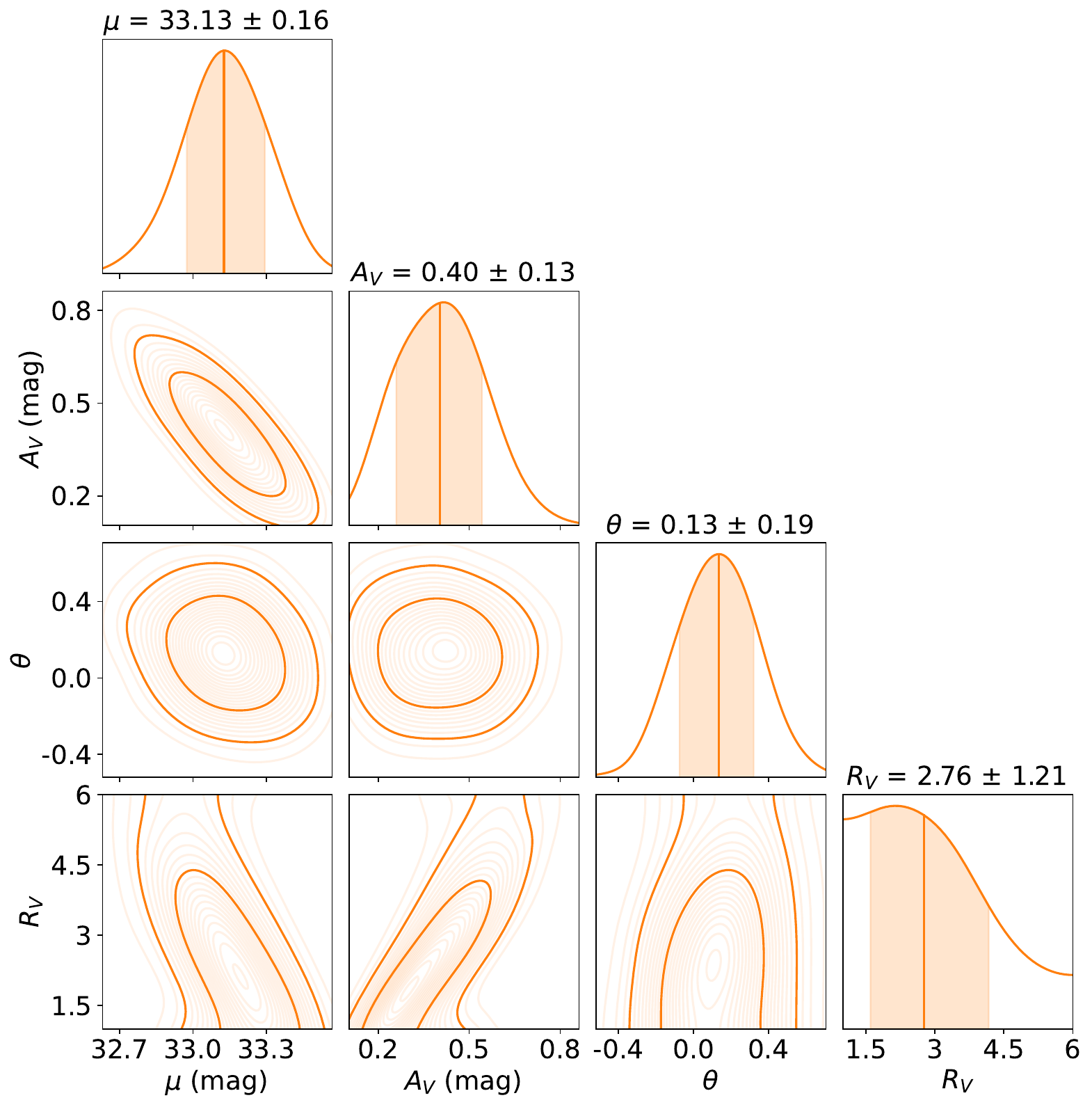}{0.5\textwidth}{}
          \fig{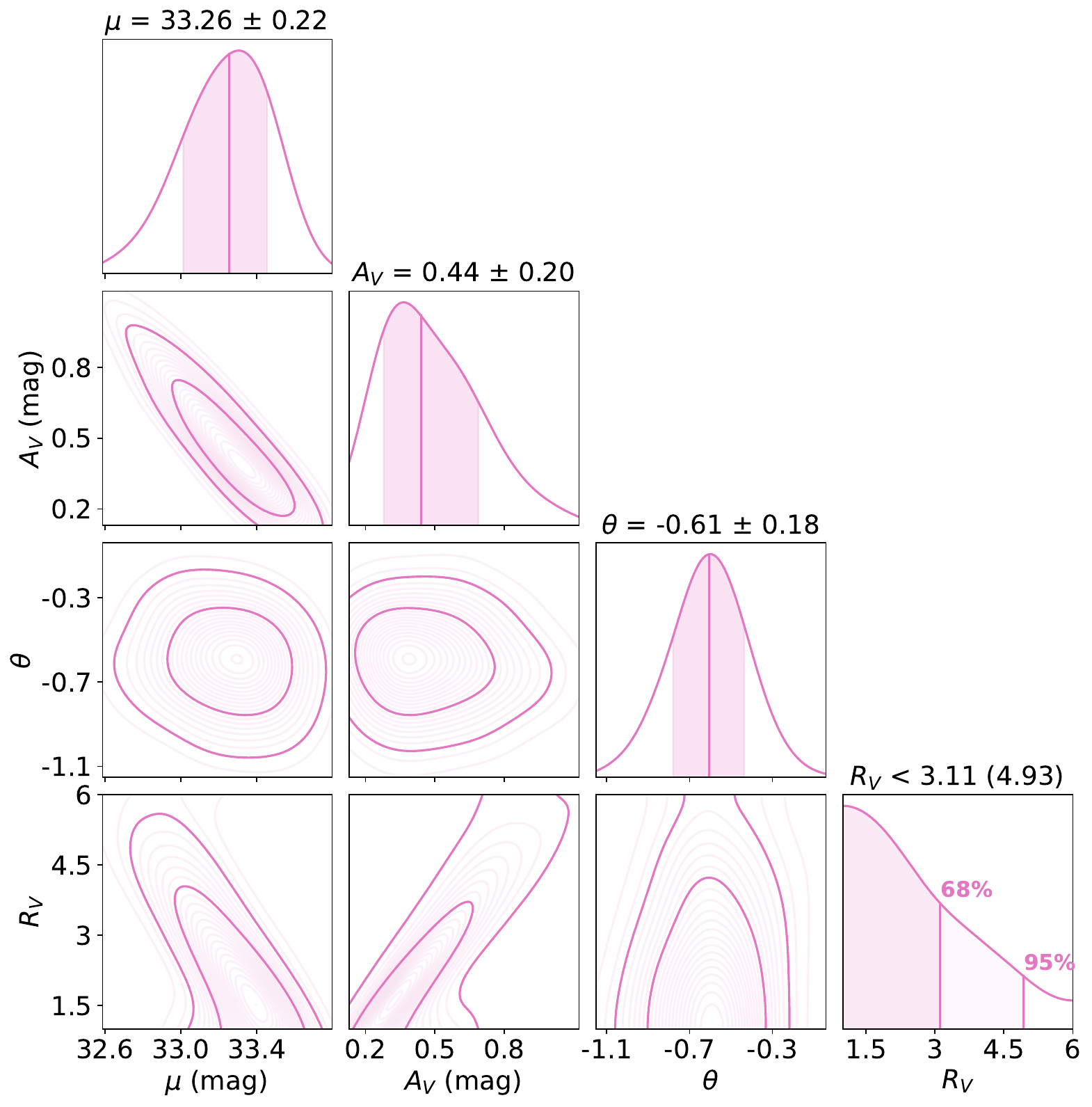}{0.5\textwidth}{}}
\vspace{-0.8cm}
\caption{Individual \textsc{BayeSN} fits to SNe~1997bq and 2008fv (as in Fig.~\ref{fig:21hprRVfit}). $R_V$ posteriors are wide owing to the moderate extinction and lack of NIR data.
}
\label{fig:97bq08fvRVfit}
\end{figure*}

\begin{deluxetable}{l c | c c c | c}
\label{tab:RVfixedfits}
\tablecaption{Posterior summaries of SN parameters from individual fits with $R_V^*=2.659$ (i.e. $R_V^s$ fixed at the W22 training value).}
\tablehead{Dataset & Bands & $\mu$ (mag)& $A^s_{V}$ (mag)&$\theta_s$
& $\hat{\sigma}_{\rm{fit},s}$(mag)\,\tablenotemark{a}}
\startdata
21hpr & $grizy$
&
$33.142\pm0.109$
 & 
$0.265\pm0.043$
 & 
 $-0.572\pm0.096$
 &  0.053\\
97bq & $BVRI$
&
$33.135\pm0.115$
 & 
$0.402\pm0.070$
 & 
$\,\,\,\,0.129\pm0.191$
& 0.065
\\
08fv & $BVR$
&
$33.224\pm0.117$
 & 
$0.499\pm0.074$
 & 
$-0.628\pm0.167$
& 0.070
\\
\enddata
\vspace{-0.1cm}
\tablenotetext{a}{The fitting uncertainties, or `measurement errors', on the individual siblings distance estimates, computed using Eq.~\ref{eq:estimatesigmatlam}.}
\end{deluxetable}

\section{$H_0$ Methods \& Additional Results}
\label{S:AppendixH0Methods}

In the upper Hubble flow rung, we use distances to a high-stellar-mass subsample of 109 SNe Ia from the Foundation DR1 sample~\citep{Foley18:found, Jones19}; this minimises the systematic contribution from the mass step towards the $H_0$ inference (because NGC 3147 is also high mass). We use W22 \textsc{BayeSN} $R_V^*=2.659$ photometric distance estimates, and redshift-based distances obtained from estimates of cosmological redshift and adopting our fiducial cosmology from \cite{Riess16}: ($\Omega_M, \Omega_{\Lambda}, H_0$) = (0.28, 0.72, 73.24 $\text{km\,s}^{-1}\text{\,Mpc}^{-1}$). We use redshifts from the NASA/IPAC Extragalactic Database (NED), corrected to the CMB frame using the flow model of \citet{Carrick15}\footnote{\url{https://cosmicflows.iap.fr/}}. The NGC 3147 Cepheid distance from \cite{Riess22} comes from their `fit 1 Baseline' analysis that uses optical+NIR `reddening-free' Wesenheit magnitudes. We use the distance derived without the inclusion of any SNe Ia in any host (as in table 6; \citealt{Riess22}). We add 0.025~mag in quadrature to the Cepheid distance uncertainty to account for the geometric distance error. The literature Cepheid distance is $\hat{\mu}^{\rm{Ceph}}_{\rm{NGC\,3147}}=33.014\pm0.167$~mag

We sample a $5\log_{10} H_0$ parameter, and transform the posterior samples to $H_0$. We marginalize also over the SN~Ia total intrinsic scatter, $\sigma_0$, and the peculiar velocity dispersion, $\sigma_{\rm{pec}}$, which is primarily constrained by the set of Hubble flow \textsc{BayeSN} distance estimates, and the redshift-based distances. The calibrator galaxy's true-distance parameter, $\mu_{\rm{NGC\,3147}}$, has an uninformative prior. We also marginalize over $\Delta M_0$, which is the difference between the calibrated absolute magnitude constant of the SNe, and the fiducial value fixed during model training; without the calibrator galaxy, this parameter is degenerate with $H_0$. Our priors are:
\begin{eqnarray}
\log_{10} H_0 &\sim& U(\log_{10} 50, \log_{10} 100)\\
\sigma_0 &\sim& U(0,1)\\
\sigma_{\rm{pec}}/c &\sim& U(0,1) \\
\mu_{\rm{NGC\,3147}} &\sim& U(-\infty, +\infty)\\
\Delta M_0 &\sim& U(-2,2).
\end{eqnarray}

In Table~\ref{tab:H0constraints}, we show $H_0$ inferences are insensitive to various modeling assumptions in the calibrator galaxy. The default configuration is to simultaneously fit the siblings' \textit{light curves} with the Cepheid and Hubble flow distances (see `Siblings Light Curves' block in Table~\ref{tab:H0constraints}). To demonstrate that these methodologies for siblings can be applied to distances obtained with any SN model, we perform additional $H_0$ inferences using only one of the individual siblings \textit{distance estimates} (\S\ref{S:fittingprocedures}; `Individual Fit Distance'), or all individual siblings distance estimates at the same time (\S\ref{S:ModellingAssumptions}; `All Individual Fit Distances'). These 2-step inferences are faster to compute, so we test the sensitivity of $H_0$ estimates to the $R_V^s$ assumptions, and the intrinsic scatter modeling assumptions, in the calibrator galaxy.

\section{Simulations}
\label{S:Simulations}

\begin{figure*}
    \includegraphics[width=0.5\linewidth]{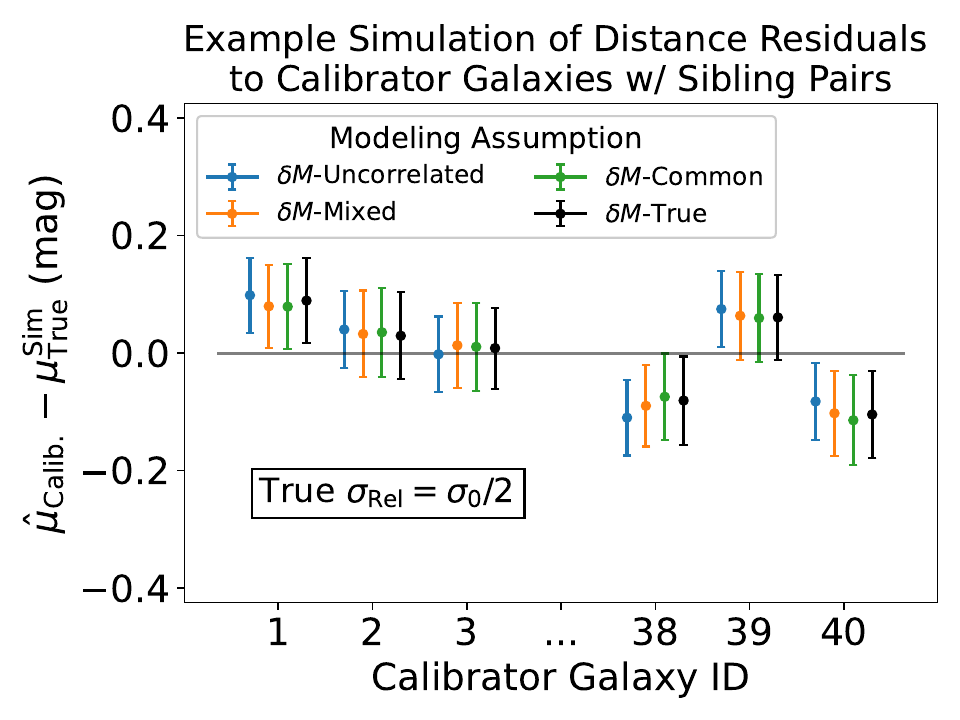}
    \includegraphics[width=0.5\linewidth]{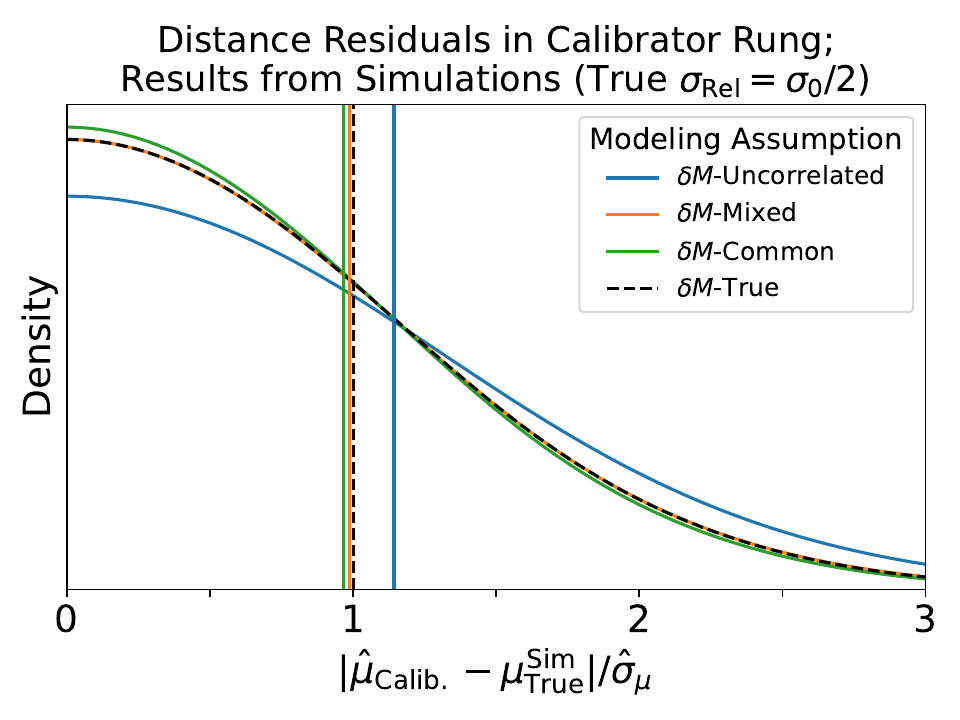}
    \caption{
    (left panel)
    Example simulation showing the distance residuals to 40 calibrator galaxies with sibling pairs, where the true $\sigma_{\rm{Rel}} = \sigma_0/2 = 0.05$~mag. The distance residual, $\hat{\mu}_{\rm{Calib.}}-\mu^{\rm{Sim}}_{\rm{True}}$, compares the true distance in the simulation, $\mu^{\rm{Sim}}_{\rm{True}}$, with the 
    calibrated SN siblings common distance estimate to the calibrator galaxy, $\hat{\mu}_{\rm{Calib.}}$. (right panel) We plot a KDE of the 40 distance residuals, $|\hat{\mu}_{\rm{Calib.}}-\mu^{\rm{Sim}}_{\rm{True}}|/\hat{\sigma}_{\mu}$, averaged over 100 simulations, which shows how well estimated the calibrator galaxy distance uncertainties are. For each galaxy, we take the posterior median distance estimate, and divide by the posterior standard deviation, $\hat{\sigma}_{\mu}$; the vertical lines show the 68\% quantiles, which should be close to 1 if the uncertainties are well estimated. The uncertainties are overestimated when adopting the $\delta M$-Common assumption (where $\sigma_{\rm{Rel}}=0$ is assumed), whereas the $\delta M$-Uncorrelated uncertainties are underestimated (where $\sigma_{\rm{Rel}}=\sigma_0$ is assumed). The $\delta M$-Mixed results (where $\sigma_{\rm{Rel}}$ is marginalized over) and the $\delta M$-True results (where $\sigma_{\rm{Rel}}$ is known and fixed in the model fit) agree with one another; moreover, the 68\% quantiles are strongly consistent with 1, indicating the distance uncertainties are well-calibrated. 
    }
    \label{fig:Calibrator}
\end{figure*}

We present here two sets of simulations to show how the siblings' intrinsic scatter modeling assumptions affect the common distance uncertainties to a single siblings galaxy, and in turn the $H_0$ uncertainties in the distance ladder. We show that adopting the $\delta M$-Uncorrelated assumption returns underestimated distance uncertainties, and hence $H_0$ uncertainties, compared to those obtained when the true $\sigma_{\rm{Rel}}$ is known (provided the true $\sigma_{\rm{Rel}}<\sigma_0$).

Firstly, we simulate three individual siblings distance estimates under a true distance $\mu=30$~mag, with individual fitting uncertainties $\hat{\sigma}_{\rm{fit}}=0.06$~mag (the median in the W22 sample), a total intrinsic scatter $\sigma_0=0.1$~mag, and various choices of relative intrinsic scatter: $\sigma_{\rm{Rel}} = (0,0.25,0.5,0.75,1)\times\sigma_0$. We then fit for the common distance, with $\sigma_0=0.1$~mag, and under the three intrinsic scatter modeling assumptions: $\delta M$-Uncorrelated ($\sigma_{\rm{Rel}}=\sigma_0$), $\delta M$-Mixed (where $\sigma_{\rm{Rel}}$ is marginalized over using a uniform prior $\sigma_{\rm{Rel}}\sim U(0,\sigma_0)$) and $\delta M$-Common ($\sigma_{\rm{Rel}}=0$). We perform 100 simulations, then compare against $\delta M$-True inferences, when the true value of $\sigma_{\rm{Rel}}$ is known in each fit.

Table~\ref{tab:musibs} shows the $\delta M$-Uncorrelated uncertainties are always smaller than the $\delta M$-True uncertainties when $\sigma_{\rm{Rel}}<\sigma_0$. Therefore, adopting the $\delta M$-Uncorrelated assumption will always underestimate the common distance uncertainty, unless the siblings are in the extreme regime of being perfectly uncorrelated: $\sigma_{\rm{Rel}}=\sigma_0$. The distance uncertainties increase as the siblings are assumed to be more correlated, and the most conservative estimates are obtained by adopting the $\delta M$-Common assumption (i.e. $\sigma_{\rm{Rel}}=0$). In the rows where $\sigma_{\rm{Rel}}\lesssim \sigma_0/2$, the $\delta M$-Mixed uncertainties are slightly underestimated, $\sim\mathcal{O}$(1~mmag), compared to the $\delta M$-True uncertainties, which is reflective of the weak constraining power on $\sigma_{\rm{Rel}}$ in the small sample size limit. However,
these underestimates are an order of magnitude smaller than the $\approx0.02-0.04$~mag underestimates from adopting the $\delta M$-Uncorrelated assumption. The $\delta M$-Mixed uncertainties match or exceed the $\delta M$-True uncertainties for $\sigma_{\rm{Rel}}\gtrsim \sigma_0/2$. Therefore, the most important modeling choice in the small sample size limit, which has the largest impact on siblings inferences, is whether to adopt the $\delta M$-Uncorrelated assumption, or either of the $\delta M$-Mixed/Common assumptions.

We further explore the effect of relative intrinsic scatter on inferences of $H_0$, by assuming calibrator galaxies in the 2nd rung of the distance ladder contain siblings. We simulate 40 calibrator galaxies with sibling pairs, and Cepheid distances with 0.1~mag measurement errors. In the 3rd rung we simulate 100 single-SN Hubble flow galaxies in the redshift range $z_{\rm{CMB}} \sim U(0.01,0.1)$, with $\sigma_{\rm{pec}}=150\,\text{km\,s}^{-1}\text{\,Mpc}^{-1}$, and heliocentric redshift measurement errors of 0.0005. The simulated true Hubble constant is $H_0=70\,\text{km\,s}^{-1}\text{\,Mpc}^{-1}$. As above, we assess recovery of $H_0$ for different intrinsic scatter modeling assumptions. Table~\ref{tab:H0sims} shows the $H_0$ uncertainties are underestimated with the $\delta M$-Uncorrelated assumption, in all cases except for when the true $\sigma_{\rm{Rel}}=\sigma_0$. With 40 sibling pairs, $\sigma_{\rm{Rel}}$ is constrained well, to the extent that the $\delta M$-Mixed uncertainties agree strongly with the $\delta M$-True uncertainties for all $\sigma_{\rm{Rel}}$ values.

To better intuit these simulations, we plot the distance residuals and their uncertainties
in Fig.~\ref{fig:Calibrator}. The left panel shows an example simulation, where the true $\sigma_{\rm{Rel}}=\sigma_0/2$. The right panel shows a KDE of the 40 distance residuals, normalized by their uncertainties, averaged over 100 simulations. With well-calibrated uncertainties, the distance residuals should be $\approx 1\sigma$ consistent with zero in 68\% of simulations; however, the residuals are consistent with zero in less than 68\% of simulations if the uncertainties are underestimated, and vice versa. The right panel shows the $\delta M$-Common, -Mixed, and -Uncorrelated assumptions lead to overestimated, well-calibrated, and underestimated uncertainties, respectively.

\begin{deluxetable}{c c c c}
\label{tab:H0constraints}
\tablecaption{Hubble constant constraints, using the siblings trio, a Cepheid distance~\cite{Riess22}, and a Hubble flow sample of 109 SNe Ia in high-stellar-mass host galaxies from Foundation DR1~\citep{Foley18:found, Jones19}.}
\tablehead{
SN Calibrator(s)&$\delta M$ Modeling Assumption\,\tablenotemark{a}&
\multicolumn{2}{c}{$H_0\, (\text{km\,s}^{-1}\text{\,Mpc}^{-1})$}\\\cline{3-4}
&& $R_V^*=2.659$\,\tablenotemark{e} & $R_V^s\sim \mathcal{U}(1,6)$
}
\startdata
Individual Fit Distance\,\tablenotemark{b} \\
\tableline
 21hpr & -
 & $ 77.9 \pm 6.4 $  & $ 77.6 \pm 6.6 $ \\
97bq & -
& $ 77.8 \pm 6.6 $ &  $ 77.4 \pm 7.6 $ \\
08fv & -
&$ 80.9 \pm 6.7 $ &  $ 81.4 \pm 8.7 $ \\
\tableline
All Individual Fit Distances\,\tablenotemark{c} \\
\tableline
 (21hpr, 97bq, 08fv) & $\delta M$-Uncorrelated & $ 78.7 \pm 6.2 $ &  $ 78.0 \pm 6.4 $ \\
 (21hpr, 97bq, 08fv) & $\delta M$-Mixed & $ 78.6 \pm 6.4 $ &  $ 78.0 \pm 6.5 $  \\
 (21hpr, 97bq, 08fv) & $\delta M$-Common & $ 78.8 \pm 6.4 $ &$ 78.0 \pm 6.5 $ \\
\tableline
 Siblings Light Curves \,\tablenotemark{d} \\
 \tableline
 (21hpr, 97bq, 08fv) & $\delta M$-Mixed & $ 78.4 \pm 6.1 $& $\mathbf{ 78.4 \pm 6.5 }$\,\tablenotemark{f} \\
\enddata
\vspace{-0.1cm}
\tablenotetext{a}{Intrinsic scatter modeling assumptions in Table~\ref{tab:IntrinsicScatterModellingAssumptions}}
\vspace{-0.2cm}
\tablenotetext{b}{NGC~3147 is calibrated with an individual-fit distance estimate to one sibling.}
\vspace{-0.2cm}
\tablenotetext{c}{NGC~3147 is calibrated using all three individual fit distance estimates.}
\vspace{-0.2cm}
\tablenotetext{d}{NGC~3147 is calibrated by jointly fitting the light curves of the siblings trio, while simultaneously fitting the Cepheid and Hubble flow distances. }
\vspace{-0.2cm}
\tablenotetext{e}{Priors on $R_V$.}
\vspace{-0.2cm}
\tablenotetext{f}{Our best $H_0$ constraint.}
\end{deluxetable}

\begin{deluxetable}{c c | c c c | c}
\label{tab:musibs}
\tablecaption{Common distance uncertainties from three simulated siblings distances, averaged across 100 simulations.}
\tablehead{
\# Siblings& True $\sigma_{\rm{Rel}}$&
\multicolumn{4}{c}{$\hat{\sigma}_{\mu}$~(mag)\,\tablenotemark{a}}\\
\cline{3-6}
&& $\delta M$-Uncorrelated & $\delta M$-Mixed & $\delta M$-Common & $\delta M$-True\,\tablenotemark{b}
}
\startdata
3 & 0 & $0.067\pm0.001$ &$0.097\pm0.002$ &$0.106\pm0.001$ &$0.106\pm0.001$  \\ 
3 & $\sigma_0/4$ & $0.067\pm0.001$ &$0.097\pm0.003$ &$0.105\pm0.001$ &$0.104\pm0.001$  \\ 
3 & $\sigma_0$/2 & $0.067\pm0.001$ &$0.096\pm0.004$ &$0.106\pm0.001$ &$0.098\pm0.001$  \\ 
3 & $3\sigma_0/4$ & $0.067\pm0.001$ &$0.095\pm0.005$ &$0.106\pm0.001$ &$0.086\pm0.001$  \\ 
3 & $\sigma_0$ & $0.067\pm0.001$ &$0.092\pm0.006$ &$0.106\pm0.001$ &$0.067\pm0.001$  \\
\enddata
\vspace{-0.1cm}
\tablenotetext{a}{The median and sample standard deviation of common-distance uncertainties across 100 simulations (details in text).}
\vspace{-0.2cm}
\tablenotetext{b}{Uncertainty on the common-$\mu$ inference when the true $\sigma_{\rm{Rel}}$ is known and fixed in the model.}
\end{deluxetable}

\begin{deluxetable}{c c c| c | c c c | c}
\label{tab:H0sims}
\tablecaption{$H_0$ recovery across 100 simulations, where siblings calibrator galaxies each have a sibling pair, and $\sigma_0=0.1$~mag.}
\tablehead{
\# Siblings&\# Calibrators& True $\sigma_{\rm{Rel}}$& $\hat{\sigma}_{\rm{Rel}}$~(mag)\,\tablenotemark{a} &
\multicolumn{4}{c}{$\hat{\sigma}_{H_0}\, (\text{km\,s}^{-1}\text{\,Mpc}^{-1})$}\\
\cline{4-8}
Calibrators&&& $\delta M$-Mixed  & $\delta M$-Uncorrelated & $\delta M$-Mixed & $\delta M$-Common & $\delta M$-True\,\tablenotemark{b}
}
\startdata
40 & 40 & 0 & $<0.03(0.05)$ &$0.77\pm0.01$ &$0.85\pm0.01$ &$0.85\pm0.01$ &$0.85\pm0.01$  \\ 
40 & 40 & $\sigma_0/4$ & $0.03\pm 0.02$ &$0.77\pm0.01$ &$0.85\pm0.01$ &$0.85\pm0.01$ &$0.85\pm0.01$  \\ 
40 & 40 & $\sigma_0$/2 & $0.05\pm0.02$ &$0.77\pm0.01$ &$0.83\pm0.01$ &$0.85\pm0.01$ &$0.83\pm0.01$  \\ 
40 & 40 & $3\sigma_0/4$ & $0.07\pm0.02$&$0.77\pm0.01$ &$0.80\pm0.02$ &$0.85\pm0.01$ &$0.81\pm0.01$  \\ 
40 & 40 & $\sigma_0$ & $>0.08(0.06)$ &$0.77\pm0.01$ &$0.79\pm0.01$ &$0.85\pm0.01$ &$0.77\pm0.01$  \\
\enddata
\vspace{-0.1cm}
\tablenotetext{a}{The $\sigma_{\rm{Rel}}$ posterior summary from the $\delta M$-Mixed fit simulations. For the posteriors peaking at the lower or upper prior boundaries, the 68\% and 95\% quantiles, or the 32\% and 5\% quantiles, are quoted, respectively.}
\vspace{-0.2cm}
\tablenotetext{b}{$H_0$ Inference when $\sigma_{\rm{Rel}}$ is known and fixed in the model.}
\end{deluxetable}



\bibliography{bib}{}
\bibliographystyle{aasjournal}



\end{document}